\documentclass[aps,prd,reprint,groupedaddress,10pt,showkeys,nofootinbib]{revtex4-2}
\usepackage[driverfallback=dvipdfm]{hyperref}
\usepackage{amsmath,amssymb}
\usepackage{mathtools}
\usepackage{fancyvrb}
\allowdisplaybreaks[1]
\newcommand{\dr}{\mathrm{d}}
\usepackage{color}

\newcommand{\intn}{\mathrm{int}}

\renewcommand{\c}{\mathrm{c}}
\newcommand{\Dete}{\mathrm{Det}}

\newcommand{\Tr}{\mathrm{Tr}}
\newcommand{\tr}{\mathrm{tr}}

\newcommand{\eff}{\mathrm{eff}}

\renewcommand{\a}{\alpha}
\renewcommand{\b}{\beta}

\begin{document}

% Use the \preprint command to place your local institutional report
% number in the upper righthand corner of the title page in preprint mode.
% Multiple \preprint commands are allowed.
% Use the 'preprintnumbers' class option to override journal defaults
% to display numbers if necessary
%\preprint{}

%Title of paper
\title{Stochastic fluctuations and stability in birth-death population dynamics: \\
two-component Langevin equation in path-integral formalism}

% repeat the \author .. \affiliation  etc. as needed
% \email, \thanks, \homepage, \altaffiliation all apply to the current
% author. Explanatory text should go in the []'s, actual e-mail
% address or url should go in the {}'s for \email and \homepage.
% Please use the appropriate macro foreach each type of information

% \affiliation command applies to all authors since the last
% \affiliation command. The \affiliation command should follow the
% other information
% \affiliation can be followed by \email, \homepage, \thanks as well.
\author{Shigehiro~Yasui}
%\email[]{shigehiro.yasui@kochi-u.ac.jp}
%\email[]{yasuis@hiroshima-u.ac.jp}
\email[]{yasuis@keio.jp}
\email[]{s-yasui@nishogakusha-u.ac.jp}
\affiliation{Center of Medical Information Science, Kochi Medical School, Kochi University, Nankoku, Kochi 783-8505, Japan}
\affiliation{International Institute for Sustainability with Knotted Chiral Meta Matter (SKCM$^{2}$),
Hiroshima University, Hiroshima 739-8511, Japan}
\affiliation{Research and Education Center for Natural Sciences,
Keio University, Hiyoshi 4-1-1, Yokohama, Kanagawa 223-8521, Japan}
\affiliation{Nishogakusha University, 6-16, Sanbancho, Chiyoda, Tokyo 102-8336, Japan}
\author{Yutaka~Hatakeyama}
\email[]{hatake@kochi-u.ac.jp}
\affiliation{Center of Medical Information Science, Kochi Medical School, Kochi University, Nankoku, Kochi 783-8505, Japan}
\author{Yoshiyasu~Okuhara}
\email[]{okuharay@kochi-u.ac.jp}
%\homepage[]{Your web page}
%\thanks{}
%\altaffiliation{}
\affiliation{Center of Medical Information Science, Kochi Medical School, Kochi University, Nankoku, Kochi 783-8505, Japan}

%Collaboration name if desired (requires use of superscriptaddress
%option in \documentclass). \noaffiliation is required (may also be
%used with the \author command).
%\collaboration can be followed by \email, \homepage, \thanks as well.
%\collaboration{}
%\noaffiliation

\date{\today}

\begin{abstract}
We discuss the stochastic process of creation and annihilation of particles, i.e., the $A^{n} \rightleftarrows B$ process in which $n$ particles $A$s and one particle $B$ are transformed to each other. Considering the case that the stochastic fluctuations are dependent on the numbers of $A$ and $B$, we apply the Langevin equation for the stochastic time-evolution of the numbers of $A$ and $B$. We analyze the Langevin equation in the path-integral formalism, and show that the new driving force is generated dynamically by the stochastic fluctuations. We present that the generated driving force leads to the nontrivial stable equilibrium state. This equilibrium state is regarded as the new state of order which is induced effectively by stochastic fluctuations. We also discuss that the formation of such equilibrium state requires at least two stochastic variables in the stochastic processes.
\end{abstract}

% insert suggested keywords - APS authors don't need to do this

%\maketitle must follow title, authors, abstract, and keywords
\maketitle

%\newpage

% body of paper here - Use proper section commands
% References should be done using the \cite, \ref, and \label commands
%\section{}
% Put \label in argument of \section for cross-referencing
%\section{\label{}}
%\subsection{}
%\subsubsection{}

%\setcounter{tocdepth}{5}
%\setcounter{secnumdepth}{5}
%\tableofcontents
%\listoffigures
%\VerbatimFootnotes

The Langevin equation describes stochastic processes in various systems, and it is now used in a wide range of research subjects.
One useful method of analysis for the Langevin equation is the path-integral formalism given by Martin-Siggia-Rose-de Dominicis-Janssen (the MSRDJ formalism)~\cite{PhysRevA.8.423,dedominicis:jpa-00216466,Janssen1976}.\footnote{See also Ref.~\cite{Peliti_1985}. As related works, the second-quantization formalism was given in Refs.~\cite{Doi1_1976,Doi2_1976}.}
The path-integral formalism, in which the action is computed by functional integrals, has formal similarities to the field theory in quantum physics~\cite{Grassberger1982,Droz_1994,Lee1995,Cardy1996,cardy1996renormalisation,Bettelheim2001,Pastor-Satorras2001,Elgart2004,Elgart2006,Andreanov2006,Mobilia2007,Butler2009,Tauber_2011,Tauber_2012,Oizumi2013,Shih2014} (see, e.g., Ref.~\cite{Dickman2003,Janssen2005,Tauber_2005,cardy_cardy_falkovich_gawedzki_2008,Tauber2009,Tauber_2014,Weber_2017,Bressloff2014} for reviews).

In analysis of the Langevin equation, one of the features of the path-integral formalism is that it allows us to evaluate the expectation values of stochastic quantities by applying systematically the computational methods
 developed in the quantum field theory.
In the path-integral formalism, ``quantum fields" correspond to stochastic variables, and ``particles" are naturally defined as a result of quantization of the quantum fields.\footnote{See also Refs.~\cite{Doi1_1976,Doi2_1976,Peliti_1985} for the second quantization as the ``particle" picture in the ``quantum field" in the Langevin equation.}
The driving and fluctuation terms in the Langevin equation are regarded as the interaction for particles, and it becomes possible to analyze the Langevin equation within the
 framework of the quantum field theory.
As subjects other than physics, the path-integral formalism
 was
 applied to biology, such as 
population dynamics~\cite{PhysRevLett.86.4183,PhysRevE.79.061128,OVASKAINEN2010643}, cell dynamics~\cite{Zhang2014}, neurodynamics~\cite{PhysRevE.101.042124,PhysRevE.105.059901}, epidemiology of infectious disease~\cite{yasui2022criticality}, and so on.

In this paper, we focus on the stochastic processes as population dynamics in the Langevin equation with multiple components, and discuss the case where the amplitude of stochastic fluctuations is not constant but {\it dependent} on the population, i.e., the particle numbers.
This is called the {\it distorted} stochastic fluctuation.
It is known that, when the amplitude of stochastic fluctuations depends on the particle numbers, 
 the average values
 of the particle numbers 
 can be affected by the fluctuations through the
 stochastic
 dynamics.
This is
 called the noise-induced transition~\cite{Horsthemke_Lefever_1998,Garc_a_Ojalvo_1999,PhysRevA.20.1628,10.1143/PTP.65.828,PhysRevE.86.010106,doi:10.1143/JPSJ.77.044002,10.1371/journal.pone.0000049}.
Such stochastic systems,
 like in living cells, chemical reaction processes of biomolecules, and so on, 
can have non-trivial fluctuations because they is usually far from the equilibrium states.
This is contrasted to a thermal equilibrium where the amplitude of stochastic fluctuations is a constant number due to the
existence of
a thermal bath.

\begin{figure}
\includegraphics[keepaspectratio, scale=0.18]{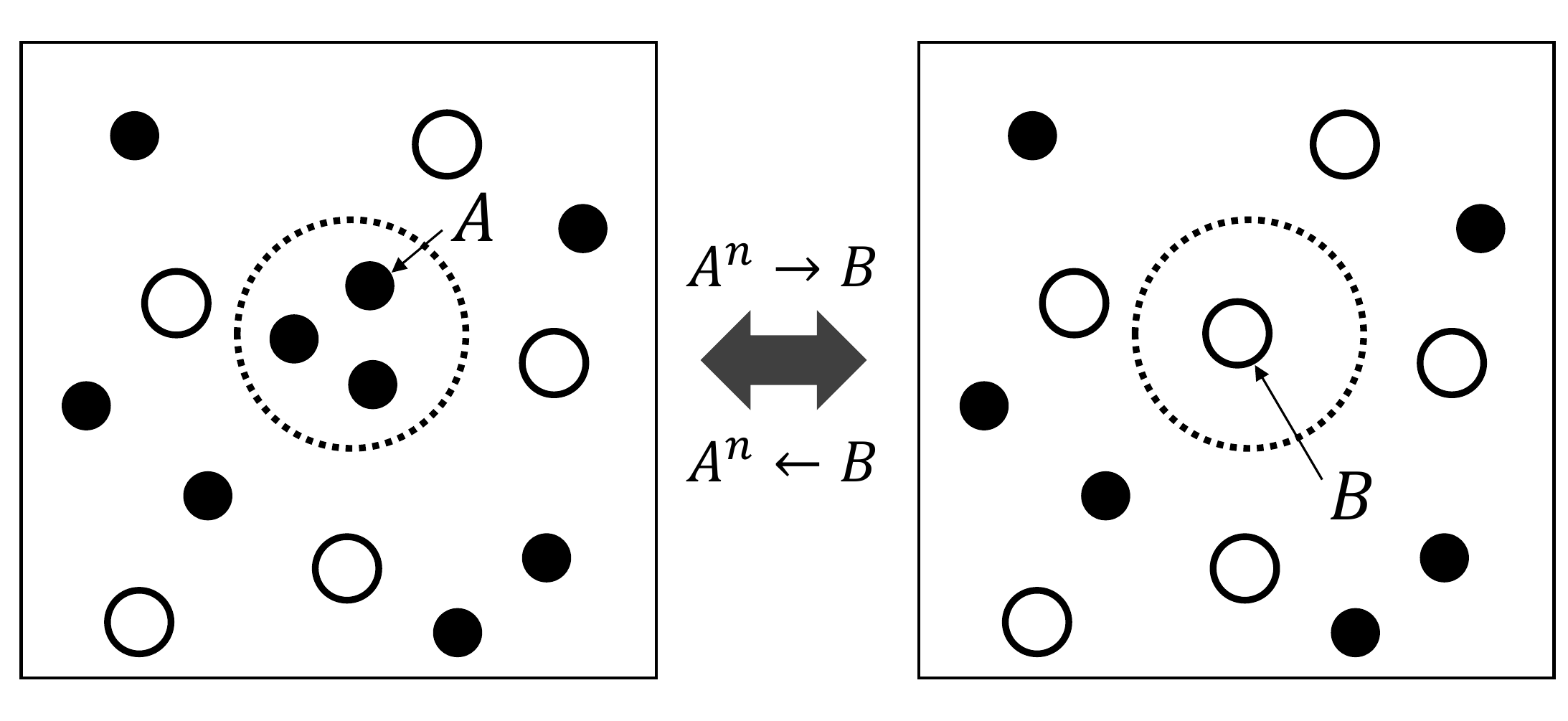}%
\caption{
The $A^{n} \rightleftarrows B$ process is shown shcematically. $A$ and $B$ are represented by black and white points, respectively.
This figure shows a snapshot that $n$ particles $A$s for $n=3$ are transformed into one particle $B$ and vice versa.
\label{fig:230417_nAB}}
\end{figure}

As a concrete example, we consider the stochastic process
 consisting of two types of particles $A$ and $B$.
 We then discuss
 the agglomeration and dispersion processes as the creation and annihilation of particles, where $n$ particles $A$s agglomerate into one particle $B$ and conversely one particle $B$ splits into $n$ particles $A$s~(Fig.~\ref{fig:230417_nAB}).
This is called the birth-death process, and it is represented symbolically by $A^{n} \rightleftarrows B$.
Here, we 
 assume that
 the agglomeration and dispersion processes are
 governed
 by
stochastic fluctuations whose amplitudes depend on the particle numbers.
In this study, thus, we focus on the distorted stochastic fluctuations stemming from the properties of the nonequilibrium state.

There are two important conclusions in the present study.
Firstly, we show that the distorted stochastic fluctuations produce new driving forces through the stochastic dynamics.
Secondly, we present that the equilibrium state can become stable when the conservation laws exist.
As for the second point, we notice that a conserved quantity in the $A^{n} \rightleftarrows B$ process is given by the sum of the number of $A$ divided by $n$ and the number of $B$, which may be denoted by $A/n+B$.
Then, we discuss that the $A^{n} \rightleftarrows B$ process can be stable (or unstable) due to such a conservation law.

This paper is organized as follows.
In Sec.~\ref{sec:formalism}, we introduce the path-integral formalism for the Langevin equation.
In Sec.~\ref{sec:perturbative_approach}, we apply the perturbative expansion for the interaction (transition) strength to obtain the time evolution of the numbers of $A$ and $B$, and estimate their average values as well as their variances in the short-time scales.
In Sec.~\ref{sec:nonperturbative_approach}, we apply the WKB approximation as a nonperturbative analysis including the higher-order terms of the interactions.
We obtain the equation-of-motion valid in the long-distance scale, and compare its solution with the result obtained in the perturbative analysis.
In Sec.~\ref{sec:discussion}, we discuss the role of multiple number of components 
  for realizing the stability of the stochastic systems.
Sec.~\ref{sec:conclusion} is devoted to our conclusion and outlook.
The details of the calculations are summarized in Appendix~\ref{sec:path_integral_supplement}.

%%%%%%%%%%%%%%%%%%%%%%%%%%%%%%%%%%%
%%%%%%%%%%%%%%%%%%%%%%%%%%%%%%%%%%%

\section{Formalism} \label{sec:formalism}
\subsection{Langevin equation} \label{sec:Langevin_equation}

We consider the $A^{n} \rightleftarrows B$ process, in which $n$ particles $A$s and one particle $B$ are transformed to each other by the driving forces accompanying the stochastic fluctuations.
We denote the numbers of $A$ and $B$ at time $\tau$ by $\a=\a(\tau)$ and $\b=\b(\tau)$, respectively, and introduce the Langevin equations for the time evolution of $\a$ and $\b$:
\begin{align}
&
   \partial_{\tau}\a + {f}(\a,\b) + \xi = 0,
\label{eq:Langevin_A}
\\ %%
&
   \partial_{\tau}\b - \frac{1}{n} {f}(\a,\b) - \frac{1}{n}\xi = 0.
\label{eq:Langevin_B}
\end{align}
In the left-hand sides,
we denote the driving force terms by ${f}(\a,\b)$ and $-{f}(\a,\b)/n$, and denote the stochastic fluctuation terms by $\xi$ and $-\xi/n$, respectively.
The stochastic variable given by $\xi=\xi(\tau)$ satisfies the following relations:
\begin{align}
   &
   \bigl\langle \xi(\tau) \bigr\rangle_{\xi} = 0,
   \label{eq:fluctuation_variable_1} \\ %%
   &
   \bigl\langle \xi(\tau) \xi(\tau') \bigr\rangle_{\xi} = 2D\bigl(\a,\b\bigr) \delta(\tau-\tau'),
\label{eq:fluctuation_variable_2}
\end{align}
where
$\langle \cdots \rangle_{\xi}$ represents the expectation value in terms of the stochastic fluctuations.\footnote{We note, in Eqs.~\eqref{eq:Langevin_A} and \eqref{eq:Langevin_B}, that the amplitude of the fluctuation term depends on $\a$ and $\beta$. As an alternative form, we may consider
\begin{align}
&
   \partial_{\tau}\a + {f}(\a,\b) + g(\a,\b) \xi = 0,
\label{eq:Langevin_A_1}
\\ %%
&
   \partial_{\tau}\b - \frac{1}{n} {f}(\a,\b) - \frac{1}{n}g(\a,\b) \xi = 0.
\label{eq:Langevin_B_1}
\end{align}
with $g(\a,\b)^{2}=2D(\a,\b)$. Here, the stochastic variable $\xi(\tau)$ is normalized by the relations,
$\displaystyle \bigl\langle \xi(\tau) \bigr\rangle_{\xi} = 0$ and $\displaystyle \bigl\langle \xi(\tau) \xi(\tau') \bigr\rangle_{\xi} = \delta(\tau-\tau')$.
The equations~\eqref{eq:Langevin_A_1} and \eqref{eq:Langevin_B_1} give an intepretation that the normalized stochastic variable, $\xi$, couples to $\a$ and $\b$ through $g(\a,\b) \xi$ and $-g(\a,\b) \xi/n$.
They are represented eventually by
the path-integral formalism same as Eq.~\eqref{eq:nAB_generating_functional}.
}
We suppose that the stochastic variable $\xi$ obeys the Gaussian (white) noise,
under the condition that the fluctuation amplitude, i.e., the correlation function of $\xi$, is not a constant value but the function of $\a$ and $\b$, as shown in Eq.~\eqref{eq:fluctuation_variable_2}.
This is the property of the distorted stochastic fluctuations.
The amplitude is denoted by $D(\a,\b)$.
The concrete form of $D(\a,\b)$ will be given later.

In the $A^{n} \rightleftarrows B$ process with Eqs.~\eqref{eq:Langevin_A} and \eqref{eq:Langevin_B}, the sum of the number of $A$ divided by $n$ and the number of $B$, i.e., $\a/n+\b$, is a constant number independent of time: $\displaystyle \partial_{\tau}\bigl(\a/n+\b\bigr)=0$.
Such a conservation is guaranteed because of the cancelations not only in the driving force terms between $A$ and $B$ but also in the fluctuation terms.\footnote{Note that this cancelation is achieved by the fact that the driving force and fluctuation terms have opposite signs as well as the scale factor $1/n$.}

In order to describe the reactions in the $A^{n} \rightleftarrows B$ process concretely, we adopt the polynomial forms for the functions, ${f}(\a,\b)$ in the driving force terms and $D(\a,\b)$ in the fluctuation terms.

As for the driving force terms, we introduce
\begin{align}
   {f}(\a,\b) = \lambda\a^{n} - \mu\b,
\label{eq:f_definition}
\end{align}
where the coefficient $\lambda$ gives the strength of the $A^{n} \rightarrow B$ transition and the coefficient $\mu$ gives the strength of the $B \rightarrow A^{n}$ transition.
This equation represents the process that $n$ particles $A$s agglomerate into one particle $B$ and one particle $B$ splits into $n$ particles $A$s.

As for the fluctuation terms, we pay a particular attention to the dependence on $\a$ and $\b$ in the amplitude $D(\a,\b)$.
We may consider that the function $D(\a,\b)$ would be a constant value independent of the numbers of $A$ and $B$,
 when the particles $A$ and $B$ are surrounded by a thermal bath,
 like in the case of Brownian motions.
 When the particles $A$ and $B$ are out of the equilibrium state,
however, the fluctuation amplitude of $A$ and $B$
 can generally be dependent on the numbers of $A$ and $B$.
We regard such a dependence as a distortion of the stochastic fluctuations.
Keeping those situations in minds, we suppose two constraint conditions for the function form of $D(\a,\b)$: (i) the stochastic fluctuation vanishes in the zero limit of the particle numbers, and (ii) the amplitude of the fluctuations is proportional to the power of the particle numbers.

Here we explain the necessity of the conditions (i) and (ii) in more details.
As for (i), we consider that the amplitude of the stochastic fluctuation in the $A^{n} \rightleftarrows B$ process becomes smaller as the numbers of $A$ and $B$ decrease, and the fluctuation vanishes eventually in the zero limit of the particle numbers.
This condition stems from the idea that the stochastic fluctuations would not exist if the particles are not present in the system.\footnote{We do not consider the outside of the system like the thermal bath, and hence there should be no source of stochastic fluctuations if there is no particle inside the system.}
As for (ii), we consider that there is a complex reaction process in shorter time scales whose microscopic mechanism should be different from the driving forces in the Langevin equation. 
In other words, we consider that the reaction process in the driving force in Eqs.~\eqref{eq:Langevin_A} and \eqref{eq:Langevin_B} is given in {\it long-time} scales, and furthermore presume that there could be higher-order correlations involving $l$ particles $A$ and $m$ particles $B$ stemming from fluctuations in {\it short-time} scales which do not appear in the time scales governed by the Langevin equations.
We regard such higher-order correlations represented by the stochastic fluctuation 
satisfying the conditions in Eqs.~\eqref{eq:fluctuation_variable_1} and \eqref{eq:fluctuation_variable_2}.

From the conditions (i) and (ii),
we consider the following power function
\begin{align}
   D(\a,\b) = \nu \a^{l} \b^{m},
\label{eq:D_definition}
\end{align}
where $\nu$ is the coefficient representing the strength of fluctuation and $l$ and $m$ are the parameters of powers.
We regard that these parameters represent the higher-order correlations among multiple number of particles.
We regard them as the phenomenological parameters independent of the driving force in the Langevin equation, because this driving force cannot determine the stochastic dynamics in short-time scales.
In the following discussions, we show that the fluctuation parameters $\nu$, $l$ and $m$ affect the time evolution of the average values and variances of the numbers of $A$ and $B$.
We furthermore present that such fluctuations provide new stable equilibrium states which cannot be achieved by the usual stochastic fluctuations possessing no dependence on the numbers of particles.

Notice that, in the present analysis, we consider a simple form of function in Eq.~\eqref{eq:D_definition}, but it can be easily extended to more complex forms for more general cases.
However, as we will discuss below, it is shown that a simple form such as Eq.~\eqref{eq:D_definition} is sufficient for us to understand the essential role of nontrivial fluctuations in stochastic processes.

Solving the Langevin equations~\eqref{eq:Langevin_A} and \eqref{eq:Langevin_B} is not an easy task in general.
Using the framework of the quantum field theory, however, we find that the stochastic fluctuation dependent on the particle numbers can be interpreted as the interactions between particles.\footnote{See, e.g., the diagrams in Figs.~\ref{fig:230511_diagram_vertex} and \eqref{fig:230512_diagram_loop}.}
Thus, the Langevin equations~\eqref{eq:Langevin_A} and \eqref{eq:Langevin_B} can be mapped to the many-body problems in quantum physics.
In the next section, we discuss how the fluctuations with amplitudes dependent on the particle numbers affect the average and variance of the particle numbers, using perturbation theory and non-perturbation theory in quantum field theory.

\subsection{Path-integral formalism} \label{sec:path_integral}

The path-integral formalism is introduced to rewrite the Langevin equations~\eqref{eq:Langevin_A} and \eqref{eq:Langevin_B} in a framework of quantum field theory.
Path integrals are given by functional forms representing the integrals over many functions, i.e., the sum of the various time-evolution paths of particles.
We introduce the notations $\phi=(\a,\b)$ and $\phi^{\ast}=(\a^{\ast},\b^{\ast})$
as two-component functions with $\a$ and $\b$,
where $\phi^{\ast}$
  is an auxiliary function complement to $\phi$.

For the Langevin equations~\eqref{eq:Langevin_A} and \eqref{eq:Langevin_B}, we define the generating functional, which is one of the most fundamental functionals in path-integral form, by
\begin{align}
   Z[j^{\ast},j]
&=
   {\cal N}
   \int {\cal D}\phi^{\ast} {\cal D}\phi \,
   \exp\biggl(
       - S
      + \int_{0}^{t}\dr \tau \, (\phi^{\ast}j+\phi j^{\ast})
   \biggr),
\label{eq:nAB_generating_functional}
\end{align}
with an action $S=S[\phi^{\ast},\phi]$ and a normalization factor ${\cal N}$.
Here $S[\phi^{\ast},\phi]$ is defined by
\begin{align}
   S[\phi^{\ast},\phi]
= S_{0}[\phi^{\ast},\phi] + S_{\intn}[\phi^{\ast},\phi],
\label{eq:S_definition}
\end{align}
where, in the right-hand side, the first term is given by
\begin{align}
   S_{0}[\phi^{\ast},\phi]
&=
   \int_{0}^{t} \dr \tau \,
   \biggl(
         \a^{\ast} \frac{\partial \a}{\partial \tau}
      + \b^{\ast} \frac{\partial \b}{\partial \tau}
         \nonumber \\ & \hspace{1em} %%
      + \a^{\ast} (\a-\bar{n}_{\a0}) \delta(\tau)
      + \b^{\ast} (\b-\bar{n}_{\b0}) \delta(\tau)
   \biggr),
\label{eq:S0_definition}
\end{align}
and the second term by
\begin{align}
   S_{\intn}[\phi^{\ast},\phi]
&=
   \int_{0}^{t} \dr \tau \,
   \Biggl(
         \biggl( \a^{\ast} - \frac{1}{n}\b^{\ast} \biggr)
         {f}(\a,\b)
         \nonumber \\ & \hspace{4em} %%
       - \biggl( \a^{\ast} - \frac{1}{n}\b^{\ast} \biggr)^{2}
         D(\a,\b)
   \Biggr).
\label{eq:Sint_definition}
\end{align}
The details of the calculations to derive the action~\eqref{eq:S_definition} from Eqs.~\eqref{eq:Langevin_A} and \eqref{eq:Langevin_B} are shown in Appendix~\ref{sec:path_integral_derivation}.\footnote{The initial conditions in the action~\eqref{eq:S0_definition} can be introduced in order for that the average of numbers of particles are given by Eqs.~\eqref{eq:perturbation_average_A_definition} and \eqref{eq:perturbation_average_B_definition}. We note that the initial conditions are not considered in the derivation in Sec.~\ref{sec:path_integral_derivation} for simplicity.}
The functions, ${f}(\a,\b)$ and $D(\a,\b)$, can be general forms in Eq.~\eqref{eq:Sint_definition}.
The external source functions, $j=(j_{\a},j_{\b})$ and $j^{\ast}=(j_{\a}^{\ast},j_{\b}^{\ast})$, are arbitrary functions introduced for the convenience of the calculations.\footnote{Note that $j$ couples to $\phi^{\ast}$ and $j^{\ast}$ couples to $\phi$.}

The action $S[\phi^{\ast},\phi]$ in Eq.~\eqref{eq:S_definition} represents the time evolution of the system from the initial time ($\tau=0$) to the final time $t$ ($\tau=t$).
Its components, $S_{0}[\phi^{\ast},\phi]$ in Eq.~\eqref{eq:S0_definition} and $S_{\intn}[\phi^{\ast},\phi]$ in Eq.~\eqref{eq:Sint_definition}, have the following meanings.
In Eq.~\eqref{eq:S0_definition}, the first two derivative terms in the integrands represent the time evolutions of $\a$ and $\b$ without any interaction.
The next two terms with the Dirac $\delta$ functions represent the initial conditions for  $\a$ and $\b$, i.e., the conditions that the average values of particle numbers at $\tau=0$ are given by constant values such as $\a(0)=\bar{n}_{\a0}$ and $\b(0)=\bar{n}_{\b0}$.
In Eq.~\eqref{eq:Sint_definition},
the first term represents the driving force and the second term represents the stochastic fluctuation, as the interactions in the $A^{n} \rightleftarrows B$ process.
We note that $\a^{\ast}$ and $\b^{\ast}$ in the integrands are combined to the polynomials of $\a^{\ast}-\b^{\ast}/n$.
This makes the quantity $\a/n+\b$ conserved, as discussed in detail later.
We note also that $\phi^{\ast}$ is included up to the quadratic terms because the stochastic fluctuations
 are Gaussian type.
If we consider higher-order correlations, then we will need to include higher-order terms, i.e., the third order or higher for $\phi^{\ast}$.

\begin{figure}
\includegraphics[keepaspectratio, scale=0.3]{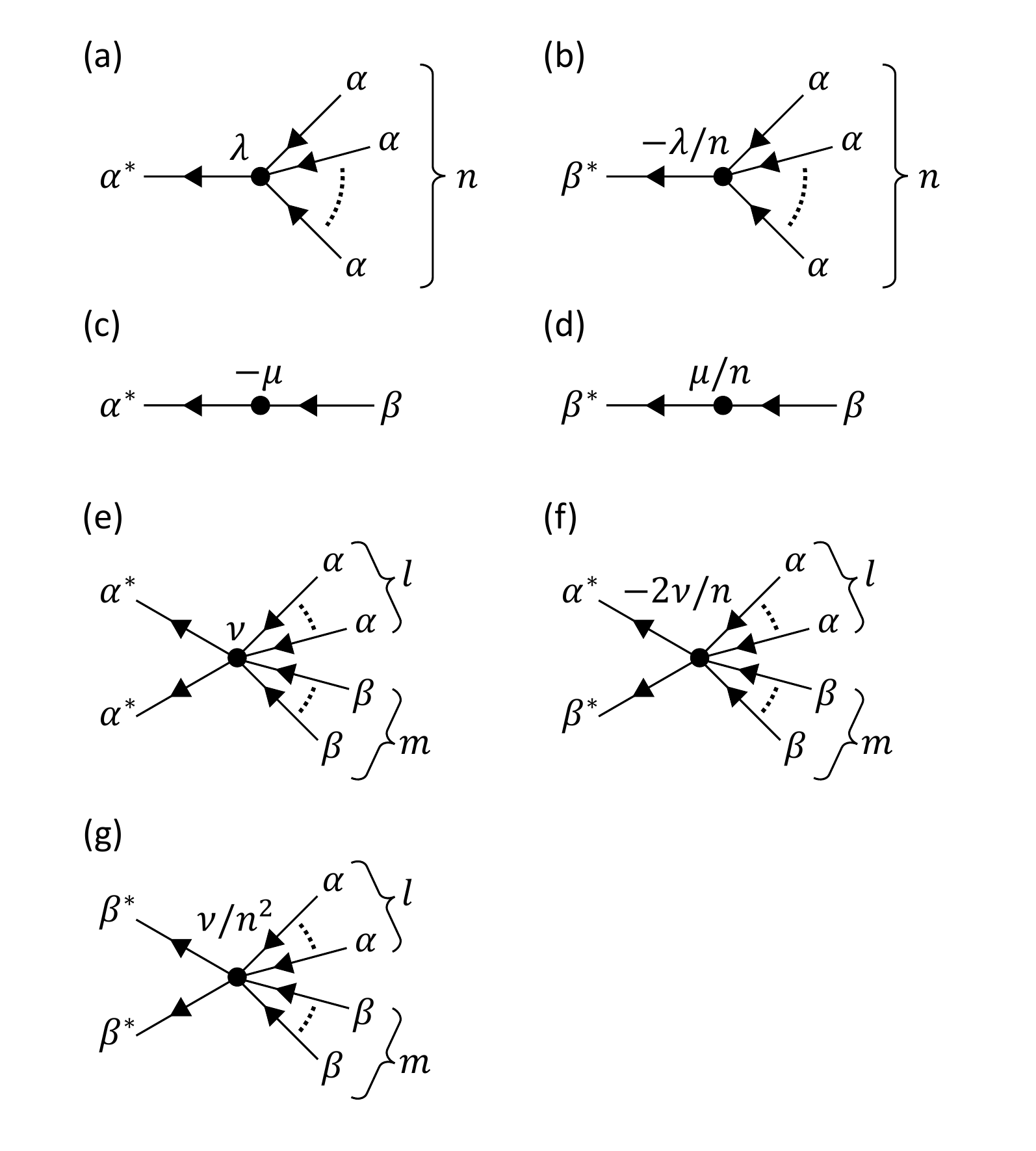}%
\caption{
The basic (tree) diagrams for the $A^{n} \rightleftarrows B$ process are shown. They are connected diagrams.
The panels (a)-(d) represent the driving forces: (a)~annihilations of $n$ particles $A$s and (b)~creation of one particle $B$ in the $A^{n} \rightarrow B$ process, and (c)~creations of $n$ particles $A$s and (d)~annihilation of one particle $B$ in the $B \rightarrow A^{n}$ process. We note the correspondence between the signs of coefficients and the creations/annihilations of particles: positive sign for annihilations of particles and negative sign for creations of particles.
The panels (e)-(g) represent the correlations by the stochastic fluctuations: (e)~fluctuations of $n$ particles $A$s, (f)~fluctuations of $n$ particles $A$s and one particle $B$, and (g)~fluctuations of one particle $B$.
\label{fig:230511_diagram_vertex}}
\end{figure}

In order to calculate the average values and variances of $\a$ and $\b$, we define another form of generating functional,
\begin{align}
   W[j^{\ast},j] = \ln Z[j^{\ast},j],
\label{eq:generating_functiona_connected}
\end{align}
instead of the original generating functional, $Z[j^{\ast},j]$, in Eq.~\eqref{eq:nAB_generating_functional}.
The generating functional $W[j^{\ast},j]$ generates the ``connected diagrams", and it removes extra information resulting from multiple counting by disconnected diagrams in $Z[j^{\ast},j]$.\footnote{The diagrams including only the connected diagrams are called connected diagrams, and play essential roles in the time evolution of the system.}
The examples of connected diagrams are shown in Figs.~\ref{fig:230511_diagram_vertex} and \ref{fig:230512_diagram_loop}.

The equation~\eqref{eq:nAB_generating_functional} is a functional integral for $\phi$ and $\phi^{\ast}$.
This formalism, called the MSRDJ theory, is similar to the quantum field theory.\footnote{More general formalism of this type of stochastic process is given by the Onsager-Machlup theory~\cite{PhysRev.91.1505}.}
It is, however, generally difficult to compute such functional integrals analytically.
To overcome this problem, in this paper, we consider approximation methods, i.e., perturbative and non-perturbative approaches.
%which are often used for systematic approximation in quantum field theory.

%%%%%%%%%%%%%%%%%%%%%%%%%%%%%%%%%%%
%%%%%%%%%%%%%%%%%%%%%%%%%%%%%%%%%%%

\section{Perturbative approach} \label{sec:perturbative_approach}

We estimate perturbatively the average numbers of $A$ and $B$ in the time evolution.
In this approach, we suppose that $\lambda$, $\mu$, and $\nu$ in the action $S[\phi^{\ast},\phi]$
 are small numbers, and hence that $S_{\intn}[\phi^{\ast},\phi]$ is treated as a perturbation against $S_{0}[\phi^{\ast},\phi]$.
By using the generating functional~\eqref{eq:generating_functiona_connected},
 we define the average numbers of $A$ and $B$ by
\begin{align}
   \langle n_{\a}(t) \rangle
= \frac{\delta W[j^{\ast},j]}{\delta j_{\a}^{\ast}(t)} \biggr|_{j=j^{\ast}=0},
\label{eq:perturbation_average_A_definition} \\ %%
   \langle n_{\b}(t) \rangle
= \frac{\delta W[j^{\ast},j]}{\delta j_{\b}^{\ast}(t)} \biggr|_{j=j^{\ast}=0},
\label{eq:perturbation_average_B_definition}
\end{align}
and the variances by
\begin{align}
   \bigl\langle \bigl( \delta n_{\a}(t) \bigr)^{2} \bigr\rangle
= \frac{\delta^{2} W[j^{\ast},j]}{\delta j_{\a}^{\ast}(t) \delta j_{\a}^{\ast}(t)} \biggr|_{j=j^{\ast}=0},
\label{eq:perturbation_variance_A_definition} \\ %%
   \bigl\langle \bigl( \delta n_{\b}(t) \bigr)^{2} \bigr\rangle
= \frac{\delta^{2} W[j^{\ast},j]}{\delta j_{\b}^{\ast}(t) \delta j_{\b}^{\ast}(t)} \biggr|_{j=j^{\ast}=0},
\label{eq:perturbation_variance_B_definition}
\end{align}
where
$\displaystyle \delta n_{\a}(t) \equiv n_{\a}(t) - \langle n_{\a}(t) \rangle$
and
$\displaystyle \delta n_{\b}(t) \equiv n_{\b}(t) - \langle n_{\b}(t) \rangle$.
We adopt approximate estimations for $\lambda$, $\mu$ and $\nu$ perturbatively by censoring the expansion series at finite order in Eqs.~\eqref{eq:perturbation_average_A_definition}, \eqref{eq:perturbation_average_B_definition}, \eqref{eq:perturbation_variance_A_definition}, and \eqref{eq:perturbation_variance_B_definition}.
The details of the calculations are shown in Appendix~\ref{sec:path_integral_perturbation}.
In the followings, we show only the results up to the first order for simplicity of our explanation.\footnote{See Eqs.~\eqref{eq:perturbation_average_A_2nd} and \eqref{eq:perturbation_average_B_2nd} for the results up to the second order.}

From Eqs.~\eqref{eq:perturbation_average_A_definition} and \eqref{eq:perturbation_average_B_definition}, we obtain the average numbers of $A$ and $B$ at the lowest order in perturbation,
\begin{align}
   \langle n_{\a}(t) \rangle
&=
   %%1
   \bar{n}_{\a0}
   %%2
 - \Biggl( %%
         \lambda \bar{n}_{\a0}^{n}
       - \mu \bar{n}_{\b0}
         \nonumber \\ & \hspace{0em} %%
      + \nu
         \biggl(
             - l \bar{n}_{\a0}^{l-1} \bar{n}_{\b0}^{m}
            + \frac{m}{n} \bar{n}_{\a0}^{l} \bar{n}_{\b0}^{m-1}
         \biggr)
   \Biggr)
   t %%%%
+ {\cal O}(t^{2}),
\label{eq:perturbation_average_A} \\
   \langle n_{\b}(t) \rangle
&=
   %%1
   \bar{n}_{\b0}
   %%2
+ \frac{1}{n}
   \Biggl( %%
         \lambda \bar{n}_{\a0}^{n}
       - \mu \bar{n}_{\b0}
         \nonumber \\ & \hspace{0em} %%
      + \nu
         \biggl(
             - l \bar{n}_{\a0}^{l-1} \bar{n}_{\b0}^{m}
            + \frac{m}{n} \bar{n}_{\a0}^{l} \bar{n}_{\b0}^{m-1}
         \biggr)
   \Biggr)
   t %%%%
+ {\cal O}(t^{2}).
\label{eq:perturbation_average_B}
\end{align}
The order of expansion for $\lambda$, $\mu$ and $\nu$ corresponds to the order of expansion for time $t$.
This correspondence arises naturally because $\lambda$, $\mu$, and $\nu$ represent the strengths of transitions per unit time in the $A^{n} \rightleftarrows B$ process.

As shown in Eqs.~\eqref{eq:perturbation_average_A} and \eqref{eq:perturbation_average_B}, the time evolutions of the numbers of $A$ and $B$ are linear in time because of the first order in the perturbation theory.
Hence, these results must be valid only in short-time scales.
We note that the absolute value of the term at the order ${\cal O}(t)$ in Eq.~\eqref{eq:perturbation_average_B} is suppressed by the factor $1/n$ in comparison with the value in Eq.~\eqref{eq:perturbation_average_A}.
This is simply because, in the Langevin equations~\eqref{eq:Langevin_A} and \eqref{eq:Langevin_B}, the absolute values of the driving force and the stochastic fluctuation for $B$ are scaled by the factor $1/n$.

\begin{figure}
\includegraphics[keepaspectratio, scale=0.3]{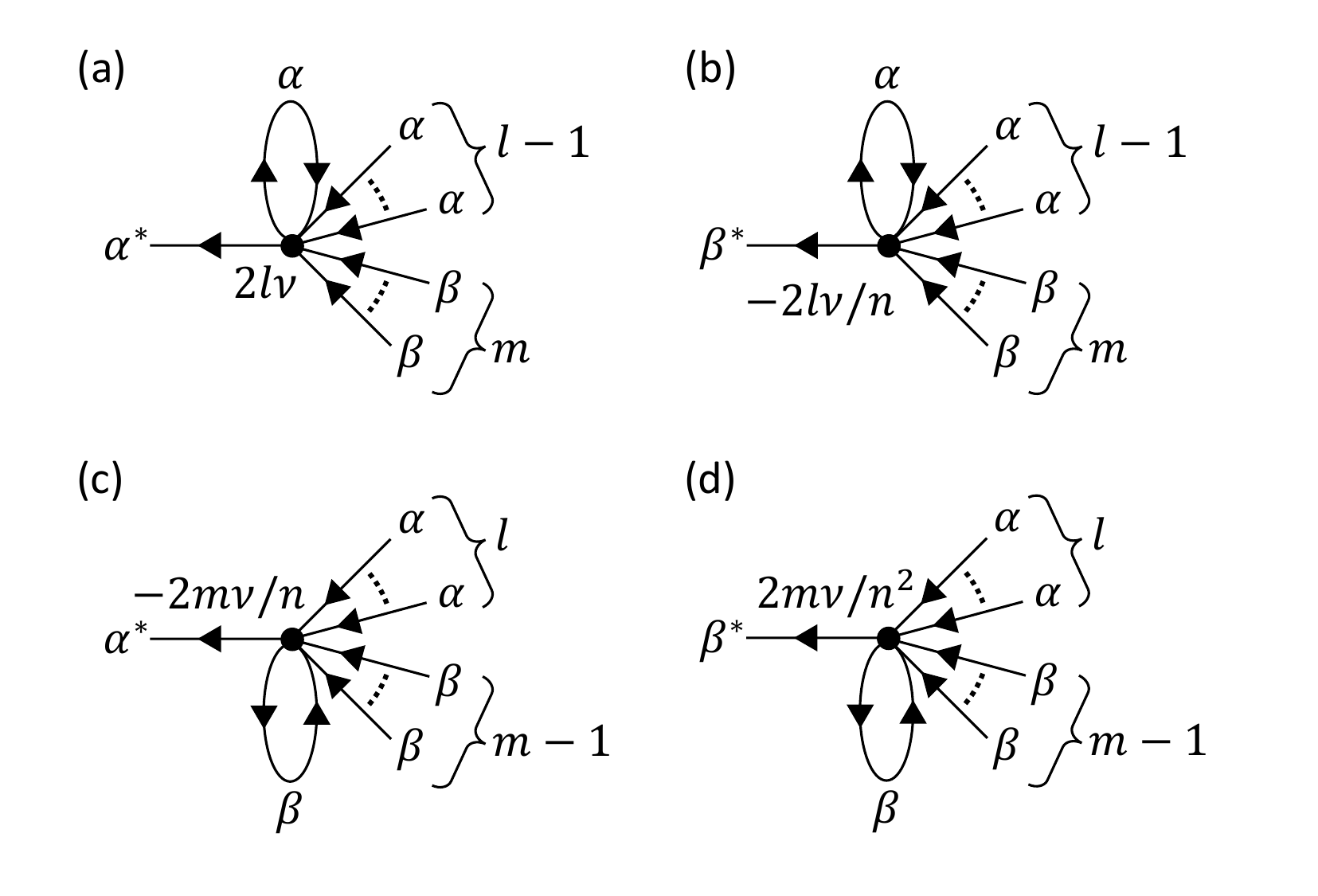}%
\caption{
The loop diagrams caused by stochastic fluctuations are shown.
It is noted that the loop diagrams are generated by the fluctuation terms shown in Figs.~\ref{fig:230511_diagram_vertex}~(e), (f), and (g).
Those diagrams are regarded as the effective driving forces which are generated dynamically by the distorted stochastic fluctuations, see the text.
\label{fig:230512_diagram_loop}}
\end{figure}

In Eqs.~\eqref{eq:perturbation_average_A} and \eqref{eq:perturbation_average_B}, we understand the roles of stochastic fluctuations in the time evolution.
For example, we focus on the behavior of $A$ in Eq.~\eqref{eq:perturbation_average_A}.
The stochastic fluctuations with the amplitude~\eqref{eq:D_definition} cause the terms, $l\nu \bar{n}_{\a0}^{l-1} \bar{n}_{\b0}^{m}t$ and $-(m\nu/n) \bar{n}_{\a0}^{l} \bar{n}_{\b0}^{m-1}t$, for the increase of $A$ in the $A^{n} \rightarrow B$ process and for the decrease of $A$ by the $B \rightarrow A^{n}$ process, respectively.\footnote{The original driving force~\eqref{eq:f_definition} causes the terms, $- \lambda \bar{n}_{\a0}^{n}t$ and $\mu \bar{n}_{\b0}t$, for the decrease of $A$ by the $A^{n} \rightarrow B$ process and for the increase of $A$ by the $B \rightarrow A^{n}$ process, respectively.}
The corresponding diagrams are shown in Fig.~\ref{fig:230512_diagram_loop}~(a), (c).
These are regarded as the result from the effective driving forces which are induced by the dependence of the power-law behavior with $l\neq0$ or $m\neq0$ in the fluctuation amplitude, see Eq.~\eqref{eq:D_definition}.
In contrast, there is no such effective driving force when the fluctuation amplitude is a constant number ($l=m=0$).
This is the case of the comventional Brownian motions.
The same discussion holds for Eq.~\eqref{eq:perturbation_average_B} (see also Fig.~\ref{fig:230512_diagram_loop}~(b), (d)).

It is important that the stochastic fluctuations lead to the dynamical stability of the system.
As for the time-evolution of $A$ in Eq.~\eqref{eq:perturbation_average_A}, we notice that increase or decrease of the number of $A$ is determined by the initial number of $A$ at $\tau=0$, i.e., $\bar{n}_{\a0}$, because the term for the increase is given by $\bar{n}_{\a0}^{l-1}$ and the term for the decrease is given by $\bar{n}_{\a0}^{l}$.\footnote{For $l>1$, for example, the number of $A$ tends to increase for a small number of $A$ because of the dominance of the term $\bar{n}_{\a0}^{l-1}$, and it tends to decrease for a large number of $A$ because of the dominance of the term $\bar{n}_{\a0}^{l}$.}
Then, we expect that the stochastic fluctuations contribute to the feedback-force mechanism leading to the stability of particle number of $A$: the number of $A$ approaches a stable equilibrium state (steady state or stationary point) due the balance among effective driving forces.
This will be shown later in numerical calculations in more details beyond the perturbation theory.
The same discussion can be applied to $B$ in Eq.~\eqref{eq:perturbation_average_B}.

%%%%%%%%%%%%%%%%%%%%%%%%%%%%%%%%%%%
%%%%%%%%%%%%%%%%%%%%%%%%%%%%%%%%%%%

From Eqs.~\eqref{eq:perturbation_average_A} and \eqref{eq:perturbation_average_B}, we confirm that a conservation law holds,
\begin{align}
   \frac{1}{n} \langle n_{\a}(t) \rangle + \langle n_{\b}(t) \rangle
= \bar{n}_{\a0} + \bar{n}_{\b0},
\end{align}
for the particle numbers of $A$ and $B$, i.e., the sum of the number of $A$ divided by $n$ and the number of $B$ is a constant number.
This conservation law should naturally hold for the higher order as well, e.g., see the second order terms shown in Eqs.~\eqref{eq:perturbation_average_A_2nd} and \eqref{eq:perturbation_average_B_2nd}.
We can indeed show directly from Eqs.~\eqref{eq:perturbation_average_A_definition} and \eqref{eq:perturbation_average_B_definition} that the conservation law is a general result independent of the order of perturbation.\footnote{Note that such a conservation law is due to the specific combination of the auxiliary functions, $\a^{\ast} - \b^{\ast}/n$, introduced in Eq.~\eqref{eq:S_definition}.}

Similarly to the discussion of the average numbers of particles,
the perturbative expansion can be applied for the variances in Eqs.~\eqref{eq:perturbation_variance_A_definition} and \eqref{eq:perturbation_variance_B_definition}.
As results, the variances of the numbers of particles for each of $A$ and $B$ in the $A^{n} \rightleftarrows B$ process are
\begin{align}
   \bigl\langle \bigl( \delta n_{\a}(t) \bigr)^{2} \bigr\rangle
&=
   \frac{1}{2!}
   2\nu
   \bar{n}_{\a}^{l} \bar{n}_{\b}^{m}
   2t
+ {\cal O}(t^{2}),
\label{eq:perturbation_variance_A}
\\ %%%%
   \bigl\langle \bigl( \delta n_{\b}(t) \bigr)^{2} \bigr\rangle
&=
   \frac{1}{2!}
   2\nu
   \bar{n}_{\a}^{l} \bar{n}_{\b}^{m}
   \frac{1}{n^{2}}
   2t
+ {\cal O}(t^{2}).
\label{eq:perturbation_variance_B}
\end{align}
From this result, we notice that the uncertainties of numbers of $A$ and $B$, i.e., the standard deviations, are proportional to $\sqrt{t}$ in short-time scales.\footnote{The standard deviation is defined as the square root of the variance.}
We also note that the variance of the number of $B$ is $1/n^{2}$ times as large as the variance of $A$.
This indicates that the standard deviation of the number of $B$ is narrowed by a factor of $1/n$ in comparison to $A$, and it can be neglected eventually in the large limit of $n$.
This result can be naturally understood because the fluctuation term in Eq.~\eqref{eq:Langevin_B} is suppressed by the factor $1/n$ against that in Eq.~\eqref{eq:Langevin_A}, and hence the stochastic fluctuation of the number of $B$ should vanish eventually in the large $n$ limit.

The variance of the combined quantity of the numbers of $A$ and $B$, $\a/n+\b$, is shown to satisfy the following relation,
\begin{align}
   \biggl\langle \biggl( \frac{1}{n} \delta n_{\a}(t)+\delta n_{\b}(t) \biggr)^{2} \biggr\rangle
= 0,
\label{eq:perturbation_variance_AB}
\end{align}
indicating the vanishing variance of $\a/n+\b$. This means that $\a/n+\b$ is strictly conserved
 in the $A^{n} \rightleftarrows B$ process, as it should be expected.
In the path-integral formalism, this conservation law is assured by the existence of the auxiliary function $\phi^{\ast}$ in Eq.~\eqref{eq:S_definition}, and hence it holds beyond the perturbative expansion in general.

As already discussed for
Eqs.~\eqref{eq:perturbation_average_A} and \eqref{eq:perturbation_average_B}, it is an interesting result that stochastic fluctuations generate new driving forces
in the $A^{n} \rightleftarrows B$ process.
It should be emphasized that, for such a phenomenon to emerge, it is essentially important that the amplitudes of the stochastic fluctuations depend on the numbers of $A$ and $B$, e.g., as shown in Eq.~\eqref{eq:D_definition}.

While we have discussed the fluctuation effects by using perturbative methods, we can discuss them by using nonperturbative methods as well, see the next section.

\section{Nonperturbative approach: WKB approximation} \label{sec:nonperturbative_approach}

As a non-perturbative analysis of the $A^{n} \rightleftarrows B$ process, we discuss an evaluation method that
considers the fluctuation effect
 around the classical solutions of $\a$ and $\b$.
This method is regarded as a non-perturbative evaluation because it includes the terms of infinite order in $\lambda$, $\mu$, and $\nu$ beyond the perturbation censoring at finite-order.\footnote{It is called the quasiclassical approximation or the Wentzel-Kramers-Brillouin (WKB) approximation in quantum physics.}

\subsection{Equation of motion} \label{sec:nonperturbative_approach_eom}

In order to estimate the numbers of $A$ and $B$ nonperturbatively,
 we consider the equations of motion for number of particles that describes the time evolution of the numbers of $A$ and $B$.
The equations of motion are defined by variational principles as
\begin{align}
   \frac{\delta \Gamma[\varphi^{\ast},\varphi]}{\delta \varphi^{\ast}} \biggr|_{\varphi^{\ast}=0} = 0,
\label{eq:eom_definition}
\end{align}
where $\Gamma[\varphi^{\ast},\varphi]$ is the effective action defined by the Legendre transformation of $-W[j,j^{\ast}]$\footnote{See Eq.~\eqref{eq:generating_functiona_connected} for $W[j,j^{\ast}]$.},
\begin{align}
   \Gamma[\varphi^{\ast},\varphi]
=
 - W[j,j^{\ast}] + j^{\ast} \varphi + j^{\ast} \varphi,
\label{eq:effective_action_definition}
\end{align}
with the functions $\varphi$ and $\varphi^{\ast}$ defined by
\begin{align}
   \varphi=\frac{\delta W}{\delta j^{\ast}}, 
   \label{eq:varphi_definition_1} \\ %%
   \varphi^{\ast}=\frac{\delta W}{\delta j}.
\label{eq:varphi_definition_2}
\end{align}
We note that the functions $\varphi$ and $\varphi^{\ast}$
represent the expectation values of $\phi$ and $\phi^{\ast}$, respectively.
Here we define $\phi_{\c}=(\a_{\c},\b_{\c})$ and $\phi_{\c}^{\ast}=(\a_{\c}^{\ast},\b_{\c}^{\ast})$ as the solutions of the
 classical equations of motion,
\begin{align}
   \frac{\delta S[\phi^{\ast},\phi]}{\delta \phi} \biggr|_{\phi_{\c}^{\ast},\phi_{\c}} - j^{\ast} &=0, \label{eq:phi_EOM1} \\ %%
   \frac{\delta S[\phi^{\ast},\phi]}{\delta \phi^{\ast}} \biggr|_{\phi_{\c}^{\ast},\phi_{\c}} - j &=0. \label{eq:phi_EOM2}
\end{align}

We may represent $\varphi$ and $\varphi^{\ast}$ by $\varphi=\phi_{\c}+\delta \phi$ and $\varphi^{\ast}=\phi_{\c}^{\ast}+\delta \phi^{\ast}$, respectively, by considering the contributions from the higher-order loops which are denoted by $\delta \phi$ and $\delta \phi^{\ast}$.
In the WKB approximation as described below, we ignore the contributions from higher-order loops, and we get the approximate relations, $\varphi \approx \phi_{\c}$ and $\varphi^{\ast} \approx \phi_{\c}^{\ast}$.
Therefore, we obtain $\Gamma[\varphi^{\ast},\varphi] \approx \Gamma[\phi_{\c}^{\ast},\phi_{\c}]$.
Note that the effect of the first order loop is included already in the effective action $\Gamma[\phi_{\c}^{\ast},\phi_{\c}]$.

The equation of motion \eqref{eq:eom_definition} represents the time evolution of the average numbers of particles $A$ and $B$.
Within the WKB approximation,
the equation of motion can be regarded as ``semiclassical'', because it gives deterministic paths of the time evolution including the fluctuation effects at the lowest order, namely in the one-loop approximation.

We use the WKB method for the path integration due to the following reasons.
Understanding how the effects of fluctuations appear in the equations of motion is a nontrivial problem,
because it is generally difficult to perform analytically the path integral of $\phi$ and $\phi^{\ast}$ in Eq.~\eqref{eq:nAB_generating_functional}.
As a trivial case, for example, let us consider $\nu=0$ in Eq.~\eqref{eq:S_definition}, i.e., the case that the action contains only the driving force term and no stochastic fluctuation term.
We then find that the effective action $\Gamma[\varphi^{\ast},\varphi]$ is simply equal to $S[\phi_{\c}^{\ast},\phi _{\c}]$,
and that the equations of motion are reduced to the simplest case that the fluctuation variables are neglected in the Langevin equations~\eqref{eq:Langevin_A} and \eqref{eq:Langevin_B} by setting $\xi=0$.
Considering $\nu \neq 0$,
in contrast, we cannot easily find the solution because the amplitude of the fluctuation variable $\xi$ depends on $\a$ and $\b$, see Eq.~\eqref{eq:D_definition}.\footnote{This can be rephrased as the difficulty of analytical computing of the path integral of the generating functional~\eqref{eq:nAB_generating_functional}. This is also regarded as the result of the nonlinear coupling for $\xi$, $\a$, and $\b$ as shown in Eqs.~\eqref{eq:Langevin_A_1} and \eqref{eq:Langevin_B_1}.}

In the WKB approximation, firstly, we separate the integral variables as $\phi=\phi_{\c}+\phi'$ and $\phi^{\ast}=\phi_{\c}^{\ast}+\phi'^{\ast}$ in
Eq.~\eqref{eq:S_definition}.
Secondly, we consider only the quadratic terms of $\phi'$ and $\phi'^{\ast}$ and perform the path integral as a Gaussian integral for them.\footnote{See Appendix~\ref{sec:path_integral_WKB} for details of the calculation.}
The functions $\phi'$ and $\phi'^{\ast}$ represent small deviations from the classical solutions, $\phi_{\c}$ and $\phi_{\c}^{\ast}$, and hence integrating for $\phi'$ and $\phi'^{\ast}$ is regarded to taking only small fluctuations around the classical paths.
In the view of the diagrams, the WKB approximation considers only one-loop diagrams as the lowest order in the loop expansion
 of the two-point propagator for the particles. Thus, it is called a one-loop approximation.\footnote{The two-point propagator represents the time evolution of the numbers of particles from past to future.}

As a result of the WKB approximation, the equations of motion \eqref{eq:eom_definition}, including stochastic fluctuations nonperturbatively at the lowest order, are obtained as
\begin{align}
   \partial_{\tau} \a_{\c}
&=
 - \lambda\a_{\c}^{n} + \mu\b_{\c}
+ \nu
   \biggl( l \a_{\c}^{l-1}\b_{\c}^{m} - \frac{m}{n}\a_{\c}^{l}\b_{\c}^{m-1} \biggr),
\label{eq:nAB2_model_WKB_A}
\\ %%
   \partial_{\tau} \b_{\c}
&=
   \frac{1}{n}
   \bigl( \lambda\a_{\c}^{n} - \mu\b_{\c} \bigr)
 - \frac{\nu}{n}
   \biggl( l \a_{\c}^{l-1}\b_{\c}^{m} - \frac{m}{n}\a_{\c}^{l}\b_{\c}^{m-1} \biggr).
\label{eq:nAB2_model_WKB_B}
\end{align}
In the right-hand sides of each equation, the terms containing $\lambda$ or $\mu$ represent the driving forces, while
 the terms containing $\nu$ represent the contributions from stochastic fluctuations.
From the latter terms, we understand
 that the stochastic fluctuations
  affect the average values of the numbers of $A$ and $B$ in the WKB approximation.
We note that the combined quantity of the numbers $A$ and $B$, $\a_{\c}/n+\b_{\c}$, is conserved in the WKB approximation, as it can be shown from Eqs.~\eqref{eq:nAB2_model_WKB_A} and \eqref{eq:nAB2_model_WKB_B}.

Similarly to the perturbation theory, we find in the WKB approximation that the driving forces are dynamically induced by the stochastic fluctuations, as shown by the terms with $\nu$ in Eqs.~\eqref{eq:nAB2_model_WKB_A} and \eqref{eq:nAB2_model_WKB_B}.
In Eq.~\eqref{eq:nAB2_model_WKB_A}, the term of $l\nu \a_{\c}^{l-1}\b_{\c}^{m}$ represents the increase of the number of $A$, and the term of $-(m\nu/n)\a_{\c}^{l}\b_{\c}^{m-1}$ represents the decrease of the number of $A$.
In Eq.~\eqref{eq:nAB2_model_WKB_B}, similarly, the term of $-(l\nu/n)\a_{\c}^{l-1}\b_{\c}^{m}$ represents the decrease of the number of $B$ and the term of $(m\nu/n^{2})\a_{\c}^{l}\b_{\c}^{m-1}$ represents the increase of the number of $B$.
These behaviors are analogously seen in the result by perturbation method, see Eqs.~\eqref{eq:perturbation_average_A} and \eqref{eq:perturbation_average_B}.

In Eqs.~\eqref{eq:nAB2_model_WKB_A} and \eqref{eq:nAB2_model_WKB_B}, we notice importantly that the realization of the effectively generated driving force requires the condition that the amplitudes of the fluctuations depend on the power of the numbers of $A$ and $B$ as parametrized by $l$ and $m$ in Eq.~\eqref{eq:D_definition}.
Indeed, we confirm that no driving force is induced by fluctuations when the amplitude of the fluctuations is a constant number independent of the numbers of $A$ and $B$, i.e., $l=m=0$.

The effective driving force can generate a new equilibrium state.
To understand this phenomena, we remember in Eqs.~\eqref{eq:nAB2_model_WKB_A} and \eqref{eq:nAB2_model_WKB_B} that there exist two different types of forces, one of which increases the numbers of $A$ and $B$ and the other decreases them, when the condition of $l\neq0$ and $m\neq0$ is satisfied.
As a result, there can appear stable equilibrium states
even in the case that the driving forces are absent and only stochastic fluctuations are present.

We find such equilibrium states by setting $\lambda=\mu=0$ and $\nu \neq 0$ in Eqs.~\eqref{eq:Langevin_A} and \eqref{eq:Langevin_B}.
Indeed, neglecting the driving force by $\lambda=\mu=0$, we find that the equations of motion \eqref{eq:nAB2_model_WKB_A} and \eqref{eq:nAB2_model_WKB_B} are reduced to
\begin{align}
   \partial_{\tau} \a_{\c}
&=
   \nu
   \biggl( l \a_{\c}^{l-1}\b_{\c}^{m} - \frac{m}{n}\a_{\c}^{l}\b_{\c}^{m-1} \biggr),
\label{eq:nAB2_model_WKB_A_fluctuation}
\\ %%
   \partial_{\tau} \b_{\c}
&=
 - \frac{\nu}{n}
   \biggl( l \a_{\c}^{l-1}\b_{\c}^{m} - \frac{m}{n}\a_{\c}^{l}\b_{\c}^{m-1} \biggr).
\label{eq:nAB2_model_WKB_B_fluctuation}
\end{align}
From Eqs.~\eqref{eq:nAB2_model_WKB_A_fluctuation} and \eqref{eq:nAB2_model_WKB_B_fluctuation}, we conclude that
the ratio of the numbers of $A$ and $B$ is determined uniquely by
\begin{align}
   \frac{\a_{\c}}{\b_{\c}} = \frac{nl}{m},
\label{eq:stable_state_ratio}
\end{align}
regardless of the value of $\nu$ and the initial conditions such as $\bar{n}_{\a0}$ and $\bar{n}_{\b0}$.\footnote{We note here that the stable equilibrium state is uniquely determined
due to the absence of the bifurcation processes in the simple form of function as shown in Eq.~\eqref{eq:D_definition}. If we would consider complicated functions for the amplitudes, we will then need to discuss bifurcation processes toward multiple different equilibrium states (cf.~Refs.~\cite{Hongler1979,Toral_2011}).}
We regard the state satisfying Eq.~\eqref{eq:stable_state_ratio} as a new state of order generated only by stochastic fluctuations.
As a comment, we note that a nontrivial equilibrium state cannot be realized for either $l=0$ or $m=0$.
This indicates that the stochastic fluctuations both of $A$ and $B$ must be simultaneously considered in order to generate a stable state induced by fluctuations.
This can be confirmed numerically as discussed in the next.

\subsection{Numerical results} \label{sec:nonperturbative_approach_numerical}

We discuss the numerical behaviors of the results in the WKB approximation.
For clarity of explanation, we set $n=2$ in the $A^{n} \rightleftarrows B$ process, and suppose the values of the model parameters as $\lambda=\mu=1$ and $\nu=0.3$.
In the present situation, we suppose that the value of $\nu$ is smaller than $\lambda$ and $\mu$ by considering that the stochastic fluctuations are more suppressed than the driving forces in the Langevin equation.
We furthermore treat $l$ and $m$ as free parameters in order to investigate the sensitivity of the results on them,
and restrict their ranges as $0 \le l \le 4$ and $0 \le m \le 3$ around $n=2$.
With this setup, we show that, even when $\nu$ is small, the stochastic fluctuations become dominant against the driving forces in the time evolution of the numbers of $A$ and $B$.

%%%%%%%%%%%%%%%%%%%%%%%%%%%%%%%%%%%
\begin{figure}
\hspace*{-2.0em}
\includegraphics[keepaspectratio, scale=0.22]{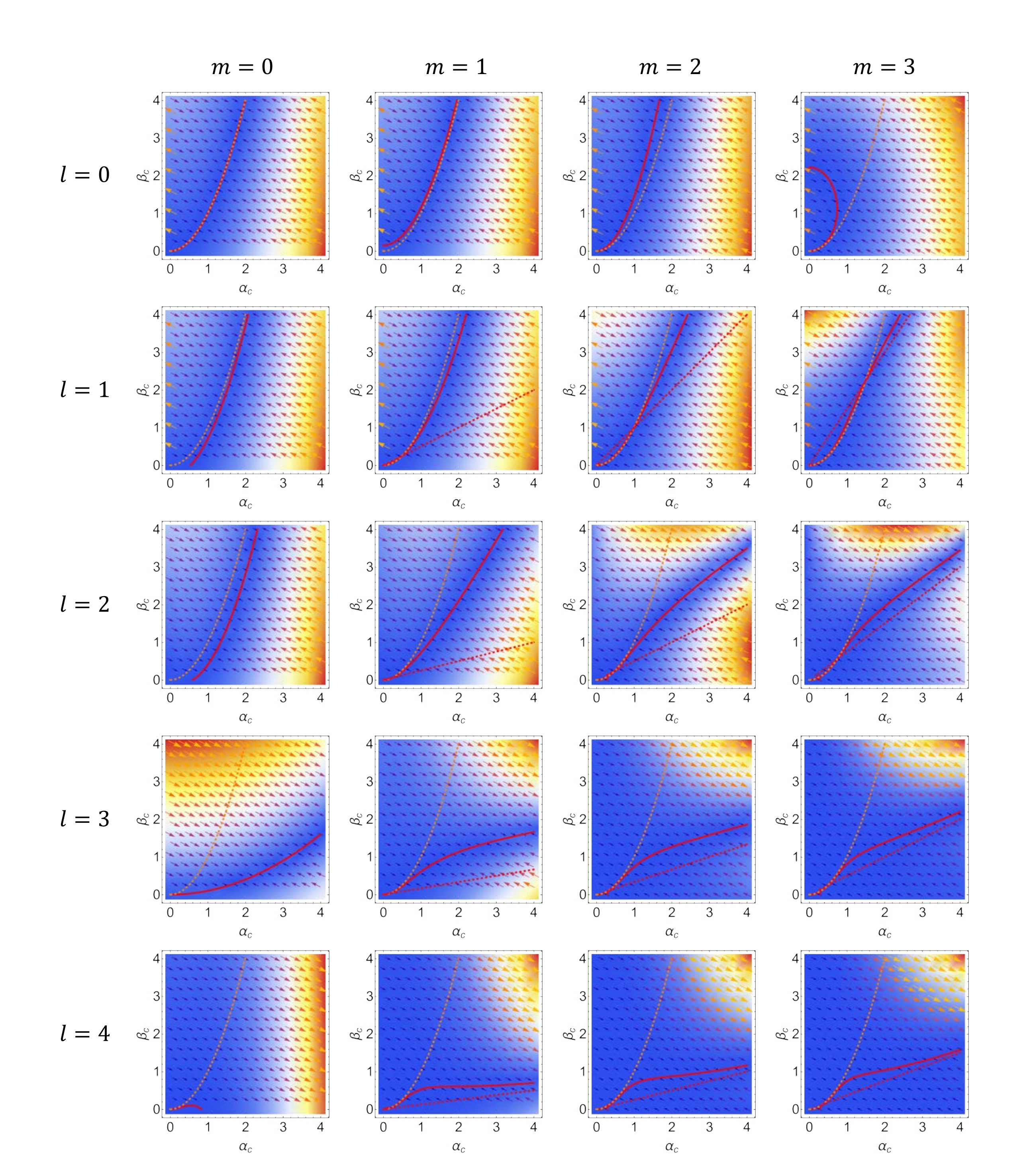}
\caption{
The flows (arrows) in the right-hand sides of Eqs.~\eqref{eq:nAB2_model_WKB_A} and \eqref{eq:nAB2_model_WKB_B} for the $A^{n} \rightleftarrows B$ process ($n=2$) are shown. The equilibrium states are indicated by lines.
The red solid lines represent the sets of the equilibrium states determined by the stationary conditions: $\partial_{\tau} \a_{\c}=\partial_{\tau} \b_{\c}=0$. 
The orange dotted lines represent the sets of the equilibrium states in the case without stochastic fluctuations ($\nu=0$), and the red dotted lines represent the ones in the case without driving forces ($\lambda=\mu=0$).
Note the existence of unstable states in some flows for $(l,m)=(4,0)$ and $(0,3)$.
\label{fig:230330_n2_l_m}}
\end{figure}
%%%%%%%%%%%%%%%%%%%%%%%%%%%%%%%%%%%

We present the numerical results in Fig.~\ref{fig:230330_n2_l_m}.
The flows (arrows) in the figure represent the right-hand sides of Eqs.~\eqref{eq:nAB2_model_WKB_A} and \eqref{eq:nAB2_model_WKB_B}, i.e., the directions for the time-evolution of states.
The red solid lines represent the sets of equilibrium states satisfying the stationary conditions: $\partial_{\tau} \a_{\c} = \partial_{\tau} \b_{\c} = 0$.
The orange dotted lines represent the sets of equilibrium states in the absence of the stochastic fluctuation terms ($\nu=0$), and the red dotted lines represent  the sets of equilibrium states in the absence of the driving force terms ($\lambda=\mu=0$).
We observe that $A$ and $B$ approach the equilibrium states along the flows at sufficiently later time.

The details of the differences for the panels of each $(l,m)$ in Fig.~\ref{fig:230330_n2_l_m} are explained as the followings.
In the simplest case of $(l,m)=(0,0)$,
the stochastic fluctuations supply no effective driving forces in Eqs.~\eqref{eq:nAB2_model_WKB_A} and \eqref{eq:nAB2_model_WKB_B}, and hence the stochastic fluctuations never affect the average values of the numbers of $A$ and $B$.
This is confirmed indeed in the figure as the red solid lines coincide with the orange dotted lines.

The case of $(l,m)=(4,3)$, on the other hand, exhibits the strongest dominance of the stochastic fluctuations.
In this case,
the stochastic fluctuations are more dominant than the driving forces for the large numbers of $A$ and $B$, while
 this relation becomes reversed for the small numbers of $A$ and $B$, as shown in Eqs.~\eqref{eq:nAB2_model_WKB_A} and \eqref{eq:nAB2_model_WKB_B}.
In fact, we confirm that the red solid lines approach the orange dotted lines for small $\a_{\c}$ and $\b_{\c}$, and, conversely,  the red solid lines approach the red dotted lines for large $\a_{\c}$ and $\b_{\c}$.

In Fig.~\ref{fig:230330_n2_l_m}, we find interestingly the behaviors of instability when only one of $l$ or $m$ is zero: either
$(l=0,m\neq0)$ or $(l\neq0,m=0)$
For example, there exist flows that never reach equilibrium states for $(l,m)=(0,3)$: they diverge at the ends of the flows.
This is also the case for $(l,m)=(4,0)$.
The reason for the existence of such instabilities can be explained: the increase and decrease of the number of $A$ or $B$ cannot be balanced only by the stochastic fluctuation terms in the absence of the original driving forces for either $(l=0,m\neq0)$ or $(l\neq0,m=0)$ in Eqs.~\eqref{eq:nAB2_model_WKB_A} and \eqref{eq:nAB2_model_WKB_B}.

As a conclusion, for the Langevin equations~\eqref{eq:Langevin_A} and \eqref{eq:Langevin_B}, we find the condition to guarantee the existence of stable equilibrium states: the amplitude of the fluctuations is constant or dependent on the number of both $A$ and $B$ ($l\neq0$ and $m\neq0$).
In the latter case, i.e., $l\neq0$ and $m\neq0$, it is especially interesting that there exists one nontrivial equilibrium state uniquely when it is generated by the stochastic fluctuations in the absence of the driving forces.

Before ending this section,
 we leave some comments on the WKB approximation used in our analysis.

Firstly, we can regard the WKB approximation as being partly relevant to a perturbative approximation.
Indeed, we can confirm that the right-hand sides of Eqs.~\eqref{eq:nAB2_model_WKB_A} and \eqref{eq:nAB2_model_WKB_B} correspond to the result from the perturbation calculation, i.e., the change of the numbers of $A$ and $B$ 
 in Eqs.~\eqref{eq:perturbation_average_A} and \eqref{eq:perturbation_average_B}.
Thus, we confirm that both are consistent to give the same results in short-time scales.
We notice, however, that the nonperturbative calculation is valid not only in short-time scales but also in long-time scales, as it is opposed to the perturbation calculation being valid only in short-time scales.
 
Secondly, the stochastic fluctuations induce, not only the effective driving forces, but also nonzero variances around the average values as shown in Eqs.~\eqref{eq:perturbation_variance_A} and \eqref{eq:perturbation_variance_B}.
In fact, the equations~\eqref{eq:perturbation_variance_A} and \eqref{eq:perturbation_variance_B} indicate that
the variances increase in the short-time scales in the time evolution of the numbers of $A$ and $B$,
leading to some uncertainties of their predictions in the later times.
Due to such uncertainties, the effective driving forces cannot be simply
regarded as the driving forces defined originally in the Langevin equation.

\section{Discussion: $A^{n} \rightleftarrows B$ v.s. $A^{n} \rightleftarrows \emptyset$}
\label{sec:discussion}

The existence of the stable equilibrium states can be roughly understood from the conservation law of $A/n+B$.
When $A$ increases as $B$ decreases, the transitions from $A$ to $B$ will occur frequently.
Conversely, when $A$ decreases as $B$ increases, the transitions from $A$ to $B$ will be suppressed. Such competing forces lead to stable equilibrium states.
We may then raise a question whether there can exist always such stable equilibrium states without the driving forces defined in the Langevin equations.
To investigate such a problem, we analyze a counter model to the $A^{n} \rightleftarrows B$ process.

For this purpose, we consider the $A^{n} \rightleftarrows \emptyset$ process
representing that $n$ particles $A$s
 vanish
 in the right direction
 and $n$ particles $A$s are created from vacuum in the left direction.
The symbol $\emptyset$ represents the vacuum state (``nothingness'').\footnote{Strictly to say, if $A$ is a real particle such as an atom or a molecule, then $A$ does not disappear to or appear from vacuum.
In the $A^{n} \rightleftarrows \emptyset$ process, we consider
 %that 
 %can be valid
  the situation that there is no need to consider the outer state where $A$ is going and from which $A$ comes into being.
  Such an absorbing and emitting state is regarded as vacuum in the present context.}
We note
that the numbers of $A$ are no longer conserved in this process.
We also suppose that the amplitude of the stochastic fluctuations in the $A^{n} \rightleftarrows \emptyset$ process depends on the number of $A$, and show that such distorted fluctuations affect the average number of $A$.
In contrast to the $A^{n} \rightleftarrows B$ process, however, we show that the $A^{n} \rightleftarrows \emptyset$ process cannot achieve stable equilibrium states when there are no driving forces but only stochastic fluctuations.

Denoting $\a(\tau)$ for the number of $A$ at time $\tau$ in the $A^{n} \rightleftarrows \emptyset$ process, we consider the Langevin equation given by
\begin{align}
   \partial_{\tau}\a + \lambda\a^{n} - \mu + \xi = 0.
\label{eq:Langevin_A_0}
\end{align}
The driving force is given by $\lambda\a^{n}-\mu$ with the constant numbers, $\lambda$ and $\mu$.
These coefficients represent the decrease and increase of the number of $A$, respectively.
The stochastic fluctuation is given by $\xi$. This is the Gaussian noise satisfying the following relations:
\begin{align}
   &
   \bigl\langle \xi(\tau) \bigr\rangle_{\xi} = 0, 
   \label{eq:fluctuation_variable_A_0_1} \\ %%
   &
   \bigl\langle \xi(\tau) \xi(\tau') \bigr\rangle_{\xi} = 2\nu \a^{l} \delta(\tau-\tau').
\label{eq:fluctuation_variable_A_0_2}
\end{align}
The parameters, $\nu$ and $l$, represent the fluctuation strength and the dependence of the fluctuation amplitude on the number of $A$, respectively.
We note that the symbols $\lambda$, $\mu$, $\nu$, $n$, and $l$ appeared already in Sec.~\ref{sec:formalism}, but we intentionally use the same symbols just to avoid the cumbersome.\footnote{Thus, they should be regarded to be different symbols in this section.}

Representing the Langevin equation~\eqref{eq:Langevin_A_0} by the path-integral formalism and adopting the WKB approximation, we obtain the equation of motion for the number of $A$,
\begin{align}
   \partial_{\tau} \a_{\c}
=
 - \lambda\a_{\c}^{n} + \mu
+ l \nu \a_{\c}^{l-1},
\label{eq:nAB2_model_WKB_A_0}
\end{align}
where the function $\a_{\c}$ denotes the ``classical path'' as discussed in Sec.~\ref{sec:nonperturbative_approach}.
The stochastic fluctuations are included at the lowest order in the loop expansion.
The term, $l \nu \a_{\c}^{l-1}$, in the right-hand side represents the effective driving force induced by the distorted stochastic fluctuations,
leading to an increase of the number of $A$.
As for the stability, however, we find that the stable equilibrium states are not guaranteed to exist in general, because their existence is much affected by our choice of the values of parameters.

In order to clarify this, as the special case, we consider the parameter set $\lambda=\mu=0$ and $\nu\neq0$ indicating the situation that there are no driving forces and only stochastic fluctuations.
Rewriting the equation~\eqref{eq:nAB2_model_WKB_A_0} as
\begin{align}
   \partial_{\tau} \a_{\c}
=
   l \nu \a_{\c}^{l-1},
\label{eq:nAB2_model_WKB_A_0_fluctuation}
\end{align}
we find no stable equilibrium state.
This is in contrast to the case in the $A^{n} \rightleftarrows B$ process.
We remember, in the $A^{n} \rightleftarrows B$ process, that there exist nontrivial stable equilibrium states due to the terms responsible for making a balance in the transitions between $A$ and $B$,
see Eqs.~\eqref{eq:nAB2_model_WKB_A_fluctuation} and \eqref{eq:nAB2_model_WKB_B_fluctuation}.
In Eq.~\eqref{eq:nAB2_model_WKB_A_0_fluctuation}, in contrast, stable equilibrium states cannot exist due to the lack of such terms.
In this way, it can be understood that the existence of the other degrees of freedom, such as $B$ in the $A^{n} \rightleftarrows B$ process, is important to realize a stable equilibrium state in the absence of driving forces.

\section{Conclusion} \label{sec:conclusion}

We analyze the Langevin equation in path-integral formalism for the $A^{n} \rightleftarrows B$ process in which $n$ particles $A$s and one particle $B$ change to each other. We focus on the distorted stochastic fluctuations: the amplitudes of the fluctuations depend on the numbers of the particles.
Based on the path-integral formalism,
we apply the perturbation theory in short-time scales and the nonperturbation theory (the WKB approximation) in long-time scales.
We conclude that the average numbers of $A$ and $B$ evolve in time not only by the driving forces but also by the stochastic fluctuations, where the distorted stochastic fluctuations induce the effective driving forces.
In particular, we emphasize that nontrivial equilibrium states are realized by the effective driving forces even in the absence of driving forces defined originally in the Langevin equation.
Such equilibrium states are considered to be dynamically-induced states of order caused by stochastic fluctuations.

As outlooks for further researches, we need to consider more realistic situations beyond the simplifications made in this study.
For example, the amplitude of the stochastic fluctuation has been supposed to be a simple power function in Eqs.~\eqref{eq:D_definition}, \eqref{eq:fluctuation_variable_A_0_1} and \eqref{eq:fluctuation_variable_A_0_2}.
However, nontrivial equilibrium states may emerge when complicated functions are adopted.
In such cases, stable equilibrium states may emerge without original driving forces even in the $A^{n} \rightleftarrows \emptyset$ process.
It would be also interesting to consider the higher-oder fluctuations beyond the Gaussian fluctuations and to study how such fluctuations can modify the lower-order fluctuations.
This analysis can be performed by the WKB approximation as a nonperturbative method.

We remember the importance of the existence of multiple numbers of particles, such as $A$ and $B$ in the $A^{n} \rightleftarrows B$ process, for the stability of the stochastic systems.
It is then an intriguing question to ask the extension of the present study to three or more particle systems.
Spatial variation of the particle numbers is also an interesting problem.
To incorporate such spatial distributions, we will need to consider the dynamics based on the Einstein-Smoluchowski equation like the Brownian motions.
Last but not least, it is also important to analyze stochastic processes by other approaches different from the Langevin equations.
For example, the analyses by the master equations reported that stochastic fluctuations give nontrivial long-time behavior in the population dynamics.\footnote{See, e.g., Ref.~\cite{PhysRevE.78.060103}. For a recent discussion of stochastic processes in population dynamics, see, for example, Refs.~\cite{OVASKAINEN2010643,doi:10.1073/pnas.1418745112} and references therein.}
Further studies for those problems are left for future challenges.

\section*{Acknowledgement}
The authors thank Chihiro~Sasaki for fruitful discussions and useful comments.
This work is supported in part by the World Premier International Research Center Initiative (WPI) under MEXT, Japan (S.~Y.).

\appendix
\section{Supplementary information for path-integral formalism} \label{sec:path_integral_supplement}

\subsection{Derivation of path-integral formalism} \label{sec:path_integral_derivation}

We show the derivation of the action~\eqref{eq:S_definition} from the Langevin equations~\eqref{eq:Langevin_A} and \eqref{eq:Langevin_B}.
First of all, we confirm that the stochastic variable $\xi$ is a function of time $\tau$, and suppose that the probability density distribution of $\xi(\tau)$ obeys the Gaussian distribution
\begin{align}
   p[\xi]
&=
   {\cal N}'
   \exp\biggl(-\int \dr \tau \, \frac{1}{4D(\a,\b)}\xi^{2}\biggr),
\end{align}
with an appropriate constant ${\cal N}'$ for normalization.
See also Eqs.~\eqref{eq:fluctuation_variable_1} and \eqref{eq:fluctuation_variable_2}.
Note that $p[\xi]$ is a functional of $\xi$.
For an arbitrary function $Q(\xi)$ of $\xi$, we obtain the expectation value
\begin{align}
   \bigl\langle Q(\xi) \bigr\rangle_{\xi}
=
   \int {\cal D} \xi \,
   Q(\xi) p[\xi],
\end{align}
by the functional integral.
For examples, setting $Q(\xi)=\xi(\tau)$ we obtain
\begin{align}
   \bigl\langle \xi(\tau) \bigr\rangle_{\xi}
=
   \int {\cal D} \xi \,
   \xi(t) p[\xi]
= 0,
\label{eq:Gaussian_expectation_1}
\end{align}
and setting $Q(\xi)=\xi(\tau)\xi(\tau')$ we obtain
\begin{align}
   \bigl\langle \xi(\tau) \xi(\tau') \bigr\rangle_{\xi}
=
   \int {\cal D}\xi \, \xi(\tau) \xi(\tau') p[\xi]
=
   2
   D(\a,\b)
   \delta(\tau-\tau').
\label{eq:Gaussian_expectation_2}
\end{align}
These are the same as Eqs.~\eqref{eq:fluctuation_variable_1} and \eqref{eq:fluctuation_variable_2}.
Thus, we understand that the function $D(\a,\b)$ represents the amplitude of the stochastic fluctuation.

We consider an arbitrary function $A(\a,\b)$ of $\a$ and $\b$
and
 give an expression of the action~\eqref{eq:S_definition} based on the Langevin equations~\eqref{eq:Langevin_A} and \eqref{eq:Langevin_B}.
First of all, we notice that the path integral for $\a$ and $\b$ is given by
\begin{align}
   A(\a_{0},\b_{0})
&=
   {\cal N}
   \int {\cal D}\a \, {\cal D}\b \, A(\a,\b)
   \delta\bigl(\a=\text{solution}\,\,\a_{0}\bigr)
   \nonumber \\ & \times
   \delta\bigl(\b=\text{solution}\,\,\b_{0}\bigr),
\end{align}
where
  $\a_{0}$ and $\b_{0}$ represent the solutions of $\a$ and $\b$ and ${\cal N}$ is a normalization factor.
Here $\delta\bigl(\a=\text{solution}\,\,\a_{0}\bigr)$ and $\delta\bigl(\b=\text{solution}\,\,\b_{0}\bigr)$ with a Dirac $\delta$ function $\delta(\cdot)$ represent that $\a$ and $\b$ are the solutions of Eqs.~\eqref{eq:Langevin_A} and \eqref{eq:Langevin_B}, which are denoted by $\a_{0}$ and $\b_{0}$, respectively.
In the following calculations, we denote $\a_{0}$ and $\b_{0}$ simply by $\a$ and $\b$, respectively, for brevity.
With this setting, we consider the expectation value of $A(\a,\b)$,
\begin{widetext}
\begin{align}
&
   \bigl\langle A(\a,\b) \bigr\rangle_{\xi}
\nonumber \\[0em] %%
&=
   \Biggl\langle
         {\cal N}
         \int {\cal D}\a \, {\cal D}\b \, A(\a,\b)
         \delta\bigl(\a=\text{solution}\bigr)
         \delta\bigl(\b=\text{solution}\bigr)
   \Biggr\rangle_{\xi}
\nonumber \\[0em] %%
&=
   {\cal N}
   \Biggl\langle
         \int {\cal D}\a \, {\cal D}\b \, A(\a,\b)
         \delta\biggl( \partial_{\tau}\a + {f}(\a,\b) + \xi \biggr)
         \delta\biggl( \partial_{\tau}\b - \frac{1}{n}{f}(\a,\b) - \frac{1}{n}\xi \biggr)
   \Biggr\rangle_{\xi}
\nonumber \\[0em] %%
&=
   {\cal N}
   \Biggl\langle
         \int {\cal D}\a \, {\cal D}\b \,
         {\cal D}\a^{\ast} \, {\cal D}\b^{\ast} \,
         A(\a,\b)
         \exp
         \Biggl(
             - \int i\a^{\ast}\biggl( \partial_{\tau}\a + {f}(\a,\b) + \xi \biggr)
         \Biggr)
         \exp
         \Biggl(
             - \int i\b^{\ast}\biggl( \partial_{\tau}\b - \frac{1}{n}{f}(\a,\b) - \frac{1}{n}\xi \biggr)
         \Biggr)
   \Biggr\rangle_{\xi}
\nonumber \\[0em] %%
&=
   {\cal N}
   \Biggl\langle
         \int {\cal D}\a \, {\cal D}\b \,
         {\cal D}\a^{\ast} \, {\cal D}\b^{\ast} \,
         A(\a,\b)
         \exp
         \Biggl(
             - \int \a^{\ast} \biggl( \partial_{\tau}\a + {f}(\a,\b) + \xi \biggr)
         \Biggr)
         \exp
         \Biggl(
             - \int \b^{\ast} \biggl( \partial_{\tau}\b - \frac{1}{n}{f}(\a,\b) - \frac{1}{n}\xi \biggr)
         \Biggr)
   \Biggr\rangle_{\xi}
\nonumber \\[0em] %%
&=
   {\cal N}
   \int {\cal D}\a \, {\cal D}\b \,
   {\cal D}\a^{\ast} \, {\cal D}\b^{\ast} \,
   A(\a,\b)
   \exp
   \Biggl(
       - \int \a^{\ast} \bigl( \partial_{\tau}\a + {f}(\a,\b) \bigr)
       - \int \b^{\ast} \biggl( \partial_{\tau}\b - \frac{1}{n}{f}(\a,\b) \biggr)
   \Biggr)
   \Biggl\langle
         \exp
         \Biggl(
             - \int \biggl( \a^{\ast} - \frac{1}{n}\b^{\ast} \biggr) \xi
         \Biggr)
   \Biggr\rangle_{\xi}
\nonumber \\[0em] %%
&=
   {\cal N}
   \int {\cal D}\a \, {\cal D}\b \,
   {\cal D}\a^{\ast} \, {\cal D}\b^{\ast} \,
   A(\a,\b)
   \exp
   \Biggl(
       - \int \a^{\ast} \bigl( \partial_{\tau}\a + {f}(\a,\b) \bigr)
       - \int \b^{\ast} \biggl( \partial_{\tau}\b - \frac{1}{n}{f}(\a,\b) \biggr)
   \Biggr)
   \exp
   \Biggl(
         \int
         D(\a,\b)
         \biggl( \a^{\ast} - \frac{1}{n}\b^{\ast} \biggr)^{2}
   \Biggr)
\nonumber \\[0em] %%
&=
   {\cal N}
   \int {\cal D}\a \, {\cal D}\b \,
   {\cal D}\a^{\ast} \, {\cal D}\b^{\ast} \,
   A(\a,\b)
   \exp
   \Biggl(
       - \int
         \Biggl(
               \a^{\ast} \partial_{\tau}\a
            + \b^{\ast} \partial_{\tau}\b
            + \biggl( \a^{\ast} - \frac{1}{n}\b^{\ast} \biggr) {f}(\a,\b)
             - \biggl( \a^{\ast} - \frac{1}{n}\b^{\ast} \biggr)^{2}
               D(\a,\b)
         \Biggr)
   \Biggr)
\nonumber \\[0em] %%
&=
   {\cal N}
   \int {\cal D}\a \, {\cal D}\b \,
   {\cal D}\a^{\ast} \, {\cal D}\b^{\ast} \,
   A(\a,\b)
   e^{- S[\phi^{\ast},\phi]}.
\label{eq:Langevin_action_calculation}
\end{align}
\end{widetext}
In the transformation of the equations, we replace the integral variables $i\a^{\ast}$ and $i\b^{\ast}$ by $\a^{\ast}$ and $\b^{\ast}$, respectively, and use the following relations as an expectation value of the function of $\xi$,
\begin{widetext}
\begin{align}
&
   \Biggl\langle
         \exp
         \Biggl(
             - \int \biggl( \a^{\ast} - \frac{1}{n}\b^{\ast} \biggr) \xi
         \Biggr)
   \Biggr\rangle_{\xi}
\nonumber \\[0em] %%
&=
   \int {\cal D}\xi \,
   \exp
   \Biggl(
       - \int \biggl( \a^{\ast} - \frac{1}{n}\b^{\ast} \biggr) \xi
   \Biggr)
   p[\xi]
\nonumber \\[0em] %%
&=
   \int {\cal D}\xi \,
   \exp
   \Biggl(
       - \int \biggl( \a^{\ast} - \frac{1}{n}\b^{\ast} \biggr) \xi
   \Biggr)
   {\cal N}'
   \exp\biggl( - \int \frac{1}{4D(\a,\b)} \xi^{2} \biggr)
\nonumber \\[0em] %%
&=
   {\cal N}'
   \int {\cal D}\xi \,
   \exp
   \Biggl(
       - \int \frac{1}{4D(\a,\b)} \Biggl( \xi^{2} + 4D(\a,\b) \biggl( \a^{\ast} - \frac{1}{n}\b^{\ast} \biggr) \xi \Biggr)
   \Biggr)
\nonumber \\[0em] %%
&=
   {\cal N}'
   \int {\cal D}\xi \,
   \exp
   \Biggl(
       - \int \frac{1}{4D(\a,\b)}
         \Biggl(
               \Biggl( \xi + 2D(\a,\b) \biggl( \a^{\ast} - \frac{1}{n}\b^{\ast} \biggr) \Biggr)^{2}
             - 4D(\a,\b)^{2} \biggl( \a^{\ast} - \frac{1}{n}\b^{\ast} \biggr)^{2}
          \Biggr)
   \Biggr)
\nonumber \\[0em] %%
&=
   {\cal N}'
   \int {\cal D}\xi \,
   \exp
   \Biggl(
       - \int \frac{1}{4D(\a,\b)}
         \Biggl( \xi + 2D(\a,\b) \biggl( \a^{\ast} - \frac{1}{n}\b^{\ast} \biggr) \Biggr)^{2}
   \Biggr)
   \exp
   \Biggl(
         \int
         D(\a,\b)
         \biggl( \a^{\ast} - \frac{1}{n}\b^{\ast} \biggr)^{2}
   \Biggr)
\nonumber \\[0em] %%
&=
   \exp
   \Biggl(
         \int
         D(\a,\b)
         \biggl( \a^{\ast} - \frac{1}{n}\b^{\ast} \biggr)^{2}
   \Biggr).
\end{align}
\end{widetext}
In the end of Eq.~\eqref{eq:Langevin_action_calculation}, we define the action $S[\phi^{\ast},\phi]$
without the terms for the initial conditions.
The equation~\eqref{eq:Langevin_action_calculation} represents that the stochastic evaluation by $\xi$ is given by the evaluation in the form of the path integral.
Thus, we have shown that the action $S[\phi^{\ast},\phi]$ in Eq.~\eqref{eq:S_definition} is naturally introduced to estimate the expectation value of $A(\a,\b)$ in the path-integral formalism.

\subsection{Calculation in perturbative approach} \label{sec:path_integral_perturbation}

We explain the details of the perturbative calculations in Sec.~\ref{sec:perturbative_approach}.
Using $\phi=(\a,\b)$, $j=(j_{\a},j_{\b})$, and $\bar{n}_{0}=(\bar{n}_{\a0},\bar{n}_{\b0})$ for simplicity of our notations, we rewrite the equation~\eqref{eq:nAB_generating_functional} in the following form:
\begin{widetext}
\begin{align}
   Z[j^{\ast},j]
&=
   {\cal N}
   \int {\cal D}\phi^{\ast} {\cal D}\phi \,
   \exp\biggl( - S_{0}[\phi^{\ast},\phi] - S_{\intn}[\phi^{\ast},\phi] + \int_{0}^{t}\dr \tau \, (\phi^{\ast}j+\phi j^{\ast}) \biggr)
\nonumber \\[0em] %%
&=
   {\cal N}
   \int {\cal D}\phi^{\ast} {\cal D}\phi \,
   \exp\Bigl( - S_{\intn}[\phi^{\ast},\phi] \Bigr)
   \exp\biggl( - S_{0}[\phi^{\ast},\phi] + \int_{0}^{t}\dr \tau \, (\phi^{\ast}j+\phi j^{\ast}) \biggr)
\nonumber \\[0em] %%
&=
   {\cal N}
   \exp\biggl( - S_{\intn}\biggl[\frac{\delta}{\delta j},\frac{\delta}{\delta j^{\ast}}\biggr] \biggr)
   Z_{0}[j^{\ast},j]
\nonumber \\[0em] %%
&=
   {\cal N}
   \exp\biggl( - S_{\intn}\biggl[\frac{\delta}{\delta j},\frac{\delta}{\delta j^{\ast}}\biggr] \biggr)
   \exp
   \biggl(
         \int_{0}^{t} \dr \tau \,
         j^{\ast}\partial_{\tau}^{-1}j
      + \bar{n}_{0} \int_{0}^{t}\dr \tau \, j^{\ast}
   \biggr)
\nonumber \\[0em] %%
&=
   {\cal N}
   G\biggl[\frac{\delta}{\delta j},\frac{\delta}{\delta j^{\ast}}\biggr]
   F\bigl[j^{\ast},j\bigr]
\nonumber \\[0em] %%
&=
   {\cal N}
   G\biggl[\frac{\delta}{\delta j},\frac{\delta}{\delta j^{\ast}}\biggr]
   F\bigl[j^{\ast},j\bigr]
   \exp
   \biggl(
         \int_{0}^{t} \dr \tau \, \bigl( \phi^{\ast}j + \phi j^{\ast} \bigr)
   \biggr)
   \biggr|_{\phi=\phi^{\ast}=0}
\nonumber \\[0em] %%
&=
   {\cal N}
   G\biggl[\frac{\delta}{\delta j},\frac{\delta}{\delta j^{\ast}}\biggr]
   F\biggl[\frac{\delta}{\delta \phi},\frac{\delta}{\delta \phi^{\ast}}\biggr]
   \exp
   \biggl(
         \int_{0}^{t} \dr \tau \, \bigl( \phi^{\ast}j + \phi j^{\ast} \bigr)
   \biggr)
   \biggr|_{\phi=\phi^{\ast}=0}
\nonumber \\[0em] %%
&=
   {\cal N}
   F\biggl[\frac{\delta}{\delta \phi},\frac{\delta}{\delta \phi^{\ast}}\biggr]
   G\biggl[\frac{\delta}{\delta j},\frac{\delta}{\delta j^{\ast}}\biggr]
   \exp
   \biggl(
         \int_{0}^{t} \dr \tau \, \bigl( \phi^{\ast}j + \phi j^{\ast} \bigr)
   \biggr)
   \biggr|_{\phi=\phi^{\ast}=0}
\nonumber \\[0em] %%
&=
   {\cal N}
   F\biggl[\frac{\delta}{\delta \phi},\frac{\delta}{\delta \phi^{\ast}}\biggr]
   G\bigl[\phi^{\ast},\phi\bigr]
   \exp
   \biggl(
         \int_{0}^{t} \dr \tau \, \bigl( \phi^{\ast}j + \phi j^{\ast} \bigr)
   \biggr)
   \biggr|_{\phi=\phi^{\ast}=0}
\nonumber \\[0em] %%
&=
   {\cal N}
   \exp
   \biggl(
         \int_{0}^{t} \dr \tau \,
         \frac{\delta}{\delta \phi}
         \partial_{\tau}^{-1}
         \frac{\delta}{\delta \phi^{\ast}}
      + \bar{n}_{0} \int_{0}^{t}\dr \tau \, j^{\ast}
   \biggr)
   \exp\Bigl( - S_{\intn}\bigl[\phi^{\ast},\phi\bigr] \Bigr)
   \exp
   \biggl(
         \int_{0}^{t} \dr \tau \, \bigl( \phi^{\ast}j + \phi j^{\ast} \bigr)
   \biggr)
   \biggr|_{\phi=\phi^{\ast}=0}
\nonumber \\[0em] %%
&=
   {\cal N}
   \exp
   \biggl(
         \int_{0}^{t} \dr \tau \,
         \frac{\delta}{\delta \phi}
         \partial_{\tau}^{-1}
         \frac{\delta}{\delta \phi^{\ast}}
      + \bar{n}_{0} \int_{0}^{t}\dr \tau \, j^{\ast}
   \biggr)
   \exp
   \biggl(
       - S_{\intn}\bigl[\phi^{\ast},\phi\bigr]
      + \int_{0}^{t} \dr \tau \, \bigl( \phi^{\ast}j + \phi j^{\ast} \bigr)
   \biggr)
   \biggr|_{\phi=\phi^{\ast}=0},
\label{eq:nAB_generating_functional_calculation}
\end{align}
\end{widetext}
where we define
\begin{align}
   Z_{0}[j^{\ast},j]
=
   {\cal N}'
   \int {\cal D}\phi^{\ast} {\cal D}\phi \,
   \exp\biggl( - S_{0} + \int_{0}^{t}\dr \tau \, (\phi^{\ast}j+\phi j^{\ast}) \biggr),
\end{align}
with the free action $S_{0}=S_{0}[\phi^{\ast},\phi]$ and the normalization factor ${\cal N}'$.
We express $Z_{0}[j^{\ast},j]$ in a compact form as
\begin{align}
   Z_{0}[j^{\ast},j]
&=
   {\cal N}'
   \exp
   \biggl(
         \int_{0}^{t} \dr \tau \,
         j^{\ast}\partial_{\tau}^{-1}j
      + \bar{n}_{0} \int_{0}^{t}\dr \tau \, j^{\ast}
   \biggr),
\end{align}
by replacing $\phi$ to $\phi+\bar{n}_{0}$ in the middle of calculation and furthermore by replacing $\phi$ and $\phi^{\ast}$ to $\phi + \partial_{\tau}^{-1}j$ and $\phi^{\ast} + \partial_{\tau}^{-1}j^{\ast}$, respectively.
In Eq.~\eqref{eq:nAB_generating_functional_calculation}, we also define
\begin{align}
   G\biggl[\frac{\delta}{\delta j},\frac{\delta}{\delta j^{\ast}}\biggr]
&=
   \exp\biggl( - S_{\intn}\biggl[\frac{\delta}{\delta j},\frac{\delta}{\delta j^{\ast}}\biggr] \biggr),
\\ %%
   F\bigl[j^{\ast},j\bigr]
&=
   \exp
   \biggl(
         \int_{0}^{t} \dr \tau \,
         j^{\ast}\partial_{\tau}^{-1}j
      + \bar{n}_{0} \int_{0}^{t}\dr \tau \, j^{\ast}
   \biggr),
\end{align}
for short equations.
Expanding the equation~\eqref{eq:nAB_generating_functional_calculation} in terms of $\lambda$, $\mu$, and $\nu$, we obtain the average values of the numbers of $A$ and $B$ from Eqs.~\eqref{eq:perturbation_average_A_definition} and \eqref{eq:perturbation_average_B_definition},
\begin{widetext}
\begin{align}
   \langle n_{\a}(t) \rangle
&=
   %%1
   \bar{n}_{\a0}
   \nonumber \\ & \hspace{0em} %%
   %%2
 - \Biggl( %%
         \lambda \bar{n}_{\a0}^{n}
       - \mu \bar{n}_{\b0}
      + \nu
         \biggl(
             - l \bar{n}_{\a0}^{l-1} \bar{n}_{\b0}^{m}
            + \frac{m}{n} \bar{n}_{\a0}^{l} \bar{n}_{\b0}^{m-1}
         \biggr)
   \Biggr)
   t %%%%
   \nonumber \\ & \hspace{0em} %%
   %%3
+ \frac{1}{2!}
   \Biggl( %%%%%%
         %%1
         2n \lambda^{2} \bar{n}_{\a0}^{2n-1}
       - 2n \lambda\mu \bar{n}_{\a0}^{n-1} \bar{n}_{\b0}
      + \frac{2}{n} \lambda\mu \bar{n}_{\a0}^{n}
       - \frac{2}{n} \mu^{2} \bar{n}_{\b0}
         \nonumber \\[0.7em] & \hspace{2em} %%
         %%2
       - 2
         \Biggl(%%%%
               \bigl( ln + n(n-1) + l(l-1) \bigr) \lambda \nu
               \bar{n}_{\a0}^{l+n-2} \bar{n}_{\b0}^{m}
            + \biggl( - m - \frac{2lm}{n} \biggr) \lambda \nu
               \bar{n}_{\a0}^{l+n-1} \bar{n}_{\b0}^{m-1}
            + \frac{1}{n^{2}} m(m-1) \lambda \nu
               \bar{n}_{\a0}^{l+n} \bar{n}_{\b0}^{m-2}
               \nonumber \\[0.7em] & \hspace{3.5em} %%%%
             - l(l-1) \mu \nu
               \bar{n}_{\a0}^{l-2} \bar{n}_{\b0}^{m+1}
            + \frac{1}{n} l(2m+1) \mu \nu
               \bar{n}_{\a0}^{l-1} \bar{n}_{\b0}^{m}
             - \frac{m^{2}}{n^{2}} \mu \nu
               \bar{n}_{\a0}^{l} \bar{n}_{\b0}^{m-1}
         \Biggr)%%%%
         \nonumber \\[0.7em] & \hspace{2em} %%
         %%3
      + l(l-1)(3l-2) \nu^{2}
         \bar{n}_{\a0}^{2l-3} \bar{n}_{\b0}^{2m}
      + \frac{1}{n^{2}} lm(15m-11) \nu^{2}
         \bar{n}_{\a0}^{2l-1} \bar{n}_{\b0}^{2m-2}
         \nonumber \\[0.7em] & \hspace{2em} %%
       - \frac{1}{n} (15l-11)lm \nu^{2}
         \bar{n}_{\a0}^{2l-2} \bar{n}_{\b0}^{2m-1}
       - \frac{1}{n^{3}} m(m-1)(5m-6) \nu^{2}
         \bar{n}_{\a0}^{2l} \bar{n}_{\b0}^{2m-3}
   \Biggr)
   \frac{1}{2}t^{2} %%%%
+ {\cal O}(t^{2}),
\label{eq:perturbation_average_A_2nd} \\
   \langle n_{\b}(t) \rangle
&=
   %%1
   \bar{n}_{\b0}
   \nonumber \\ & \hspace{0em} %%
   %%2
+ \frac{1}{n}
   \Biggl( %%
         \lambda \bar{n}_{\a0}^{n}
       - \mu \bar{n}_{\b0}
      + \nu
         \biggl(
             - l \bar{n}_{\a0}^{l-1} \bar{n}_{\b0}^{m}
            + \frac{m}{n} \bar{n}_{\a0}^{l} \bar{n}_{\b0}^{m-1}
         \biggr)
   \Biggr)
   t %%%%
   \nonumber \\ & \hspace{0em} %%
   %%3
 - \frac{1}{n}
   \frac{1}{2!}
   \Biggl( %%%%%%
         %%1
         2n \lambda^{2} \bar{n}_{\a0}^{2n-1}
         %%a102
       - 2n \lambda\mu \bar{n}_{\a0}^{n-1} \bar{n}_{\b0}
         %%a106
      + \frac{2}{n} \lambda\mu \bar{n}_{\a0}^{n}
         %\text{------(a106,b109)}
         %%a107
       - \frac{2}{n}
         \mu^{2} \bar{n}_{\b0}
         %\text{------(a107,b110)}
         \nonumber \\[0.7em] & \hspace{2em} %%
         %%2
       - 2
         \Biggl(%%%%
               \bigl( ln + n(n-1) + l(l-1) \bigr) \lambda \nu
               \bar{n}_{\a0}^{l+n-2} \bar{n}_{\b0}^{m}
               %\text{------(a01,b07,b02)} %%%%
            + \biggl( - m - \frac{2lm}{n} \biggr) \lambda \nu
               \bar{n}_{\a0}^{l+n-1} \bar{n}_{\b0}^{m-1}
               %\text{------(a02,a07,b08,b03)} %%%%
            + \frac{1}{n^{2}} m(m-1) \lambda \nu
               \bar{n}_{\a0}^{l+n} \bar{n}_{\b0}^{m-2}
               %\text{------(a08,b09)} %%%%
               \nonumber \\[0.7em] & \hspace{3.5em} %%%%
             - l(l-1) \mu \nu
               \bar{n}_{\a0}^{l-2} \bar{n}_{\b0}^{m+1}
               %\text{------(a04,b05)} %%%%
            + \frac{1}{n} l(2m+1) \mu \nu
               \bar{n}_{\a0}^{l-1} \bar{n}_{\b0}^{m}
               %\text{------(a05,a10,b11,b06)} %%%%
             - \frac{m^{2}}{n^{2}} \mu \nu
               \bar{n}_{\a0}^{l} \bar{n}_{\b0}^{m-1}
               %\text{------(a06,a11,b12)}
         \Biggr)%%%%
         \nonumber \\[0.7em] & \hspace{2em} %%
         %%3
      + l(l-1)(3l-2) \nu^{2}
         \bar{n}_{\a0}^{2l-3} \bar{n}_{\b0}^{2m}
         %\text{------(a01,b04)} %%%%
      + \frac{1}{n^{2}} lm(15m-11) \nu^{2}
         \bar{n}_{\a0}^{2l-1} \bar{n}_{\b0}^{2m-2}
         %\text{------(a02,a05,b06)} %%%%
         \nonumber \\[0.7em] & \hspace{2em} %%
       - \frac{1}{n} (15l-11)lm \nu^{2}
         \bar{n}_{\a0}^{2l-2} \bar{n}_{\b0}^{2m-1}
         %\text{------(a04,b02,b05)} %%%%
       - \frac{1}{n^{3}} m(m-1)(5m-6) \nu^{2}
         \bar{n}_{\a0}^{2l} \bar{n}_{\b0}^{2m-3}
         %\text{------(a06,b03)} %%%%
   \Biggr)
   \frac{1}{2}t^{2} %%%%
+ {\cal O}(t^{2}),
\label{eq:perturbation_average_B_2nd}
\end{align}
\end{widetext}
up to the second order perturbations.
The expressions~\eqref{eq:perturbation_average_A} and \eqref{eq:perturbation_average_B} in the main text are the results up to the first order.
Similarly, from Eqs.~\eqref{eq:perturbation_variance_A_definition} and \eqref{eq:perturbation_variance_B_definition}, we obtain the results of variances shown in Eqs.~\eqref{eq:perturbation_variance_A}, \eqref{eq:perturbation_variance_B}, and \eqref{eq:perturbation_variance_AB}.

\subsection{Calculation in WKB approximation} \label{sec:path_integral_WKB} 

We explain the details of the WKB approximation in Sec.~\ref{sec:nonperturbative_approach}.
For the action $S[\phi^{\ast},\phi]$ in Eq.~\eqref{eq:S_definition}, we suppose that $\phi_{\c}=(\a_{\c},\b_{\c})$ and $\phi_{\c}^{\ast}=(\a_{\c}^{\ast},\b_{\c}^{\ast})$ are the solutions of the equations of motion~\eqref{eq:phi_EOM1} and \eqref{eq:phi_EOM2}.
By applying the Taylor expansion in functional derivative, we obtain
\begin{align}
&
   S[\phi^{\ast},\phi] - \int \a^{\ast}j_{\a} - \int \b^{\ast}j_{\b} - \int \a j_{\a}^{\ast} - \int \b j_{\b}^{\ast}
\nonumber \\[0em] %%
&=
   S[\phi_{\c}^{\ast},\phi_{\c}] - \int \a_{\c}^{\ast}j_{\a} - \int \b_{\c}^{\ast}j_{\b} - \int \a_{\c} j_{\a}^{\ast} - \int \b_{\c} j_{\b}^{\ast}
   \nonumber \\ & %%
+ \frac{1}{2}
   \int
   \frac{\delta^{2} S}{\delta \a^{\ast} \delta \a^{\ast}} \biggr|_{\c} \bigl( \a^{\ast} - \a_{\c}^{\ast} \bigr)^{2}
   %\nonumber \\ & %%
+ \frac{1}{2}
   \int
   \frac{\delta^{2} S}{\delta \b^{\ast} \delta \b^{\ast}} \biggr|_{\c} \bigl( \b^{\ast} - \b_{\c}^{\ast} \bigr)^{2}
   \nonumber \\ & %%
+ \int
   \frac{\delta^{2} S}{\delta \a^{\ast} \delta \b^{\ast}} \biggr|_{\c} \bigl( \a^{\ast} - \a_{\c}^{\ast} \bigr) \bigl( \b^{\ast} - \b_{\c}^{\ast} \bigr)
   \nonumber \\ & %%
+ \frac{1}{2}
   \int
   \frac{\delta^{2} S}{\delta \a \delta \a} \biggr|_{\c} \bigl( \a - \a_{\c} \bigr)^{2}
   %\nonumber \\ & %%
+ \frac{1}{2}
   \int
   \frac{\delta^{2} S}{\delta \b \delta \b} \biggr|_{\c} \bigl( \b - \b_{\c} \bigr)^{2}
   \nonumber \\ & %%
+ \int
   \frac{\delta^{2} S}{\delta \a \delta \b} \biggr|_{\c} \bigl( \a - \a_{\c} \bigr) \bigl( \b - \b_{\c} \bigr)
   \nonumber \\ & %%
+ \int
   \frac{\delta^{2} S}{\delta \a^{\ast} \delta \a} \biggr|_{\c} \bigl( \a^{\ast} - \a_{\c}^{\ast} \bigr) \bigl( \a - \a_{\c} \bigr)
   \nonumber \\ & %%
+ \int
   \frac{\delta^{2} S}{\delta \a^{\ast} \delta \b} \biggr|_{\c} \bigl( \a^{\ast} - \a_{\c}^{\ast} \bigr) \bigl( \b - \b_{\c} \bigr)
   \nonumber \\ & %%
+ \int
   \frac{\delta^{2} S}{\delta \b^{\ast} \delta \a} \biggr|_{\c} \bigl( \b^{\ast} - \b_{\c}^{\ast} \bigr) \bigl( \a - \a_{\c} \bigr)
   \nonumber \\ & %%
+ \int
   \frac{\delta^{2} S}{\delta \b^{\ast} \delta \b} \biggr|_{\c} \bigl( \b^{\ast} - \b_{\c}^{\ast} \bigr) \bigl( \b - \b_{\c} \bigr)
   \nonumber \\ & %%
+ \text{higher-order}
\nonumber \\[0em] %%
&\approx
   S[\phi_{\c}^{\ast},\phi_{\c}]
   %\nonumber \\ & %%
 - \int \a_{\c}^{\ast}j_{\a} - \int \b_{\c}^{\ast}j_{\b} - \int \a_{\c} j_{\a}^{\ast} - \int \b_{\c} j_{\b}^{\ast}
   \nonumber \\ & %%
+ \int \tilde{\varphi}^{t} {\cal M}_{\c}[\phi_{\c}^{\ast},\phi_{\c}] \tilde{\varphi},
\label{eq:action_Taylor_expansion}
\end{align}
where $\displaystyle \cdots\Bigr|_{\c} = \cdots\Bigr|_{\phi_{\c}^{\ast},\phi_{\c}}$ represents the substitution of $\phi_{\c}$ and $\phi_{\c}^{\ast}$ into $\phi$ and $\phi^{\ast}$, respectively.
For the short notation, we use $\displaystyle \int \dots$ for the integral $\displaystyle \int \dr \tau \, \dots$.
In the above transformation of equations, we use the equations of motion of $\a_{\c}^{\ast}$, $\b_{\c}^{\ast}$, $\a_{\c}$, and $\b_{\c}$ defined by Eqs.~\eqref{eq:phi_EOM1} and \eqref{eq:phi_EOM2}.
In Eq.~\eqref{eq:action_Taylor_expansion}, we represent the second-order terms by
\begin{align}
   \int
   \frac{\delta^{2} S}{\delta \a^{\ast} \delta \a^{\ast}} \biggr|_{\c} \bigl( \a^{\ast} - \a_{\c}^{\ast} \bigr)^{2}
&=
   \int \dr \tau_{1} \, \dr \tau_{2}
   \frac{\delta^{2}S[\phi^{\ast},\phi]}{\delta\a^{\ast}(\tau_{1})\delta\a^{\ast}(\tau_{2})} \biggr|_{\phi_{\c}^{\ast},\phi_{\c}}
   \nonumber \\ & \times %%
   \bigl(\a^{\ast}(\tau_{1})-\a_{\c}^{\ast}(\tau_{1})\bigr)
   \nonumber \\ & \times %%
   \bigl(\a^{\ast}(\tau_{2})-\a_{\c}^{\ast}(\tau_{2})\bigr),
\nonumber \\ %%
\dots,
\end{align}
where for brevity we define the following equations,
\begin{align}
   I_{\a_{\c}^{\ast}\a_{\c}^{\ast}} &= \frac{1}{2}\frac{\delta^{2} S}{\delta \a^{\ast} \delta \a^{\ast}} \biggr|_{\c}, \\ %%
   I_{\b_{\c}^{\ast}\b_{\c}^{\ast}} &= \frac{1}{2}\frac{\delta^{2} S}{\delta \b^{\ast} \delta \b^{\ast}} \biggr|_{\c}, \\ %%
   I_{\a_{\c}^{\ast}\b_{\c}^{\ast}} &= \frac{\delta^{2} S}{\delta \a^{\ast} \delta \b^{\ast}} \biggr|_{\c}, \\ %%
   I_{\a_{\c}\a_{\c}} &= \frac{1}{2}\frac{\delta^{2} S}{\delta \a \delta \a} \biggr|_{\c}, \\ %%
   I_{\b_{\c}\b_{\c}} &= \frac{1}{2}\frac{\delta^{2} S}{\delta \b \delta \b} \biggr|_{\c}, \\ %%
   I_{\a_{\c}\b_{\c}} &= \frac{\delta^{2} S}{\delta \a \delta \b} \biggr|_{\c}, \\ %%
   I_{\a_{\c}^{\ast}\a_{\c}} &= \frac{\delta^{2} S}{\delta \a^{\ast} \delta \a} \biggr|_{\c}, \\ %%
   I_{\a_{\c}^{\ast}\b_{\c}} &= \frac{\delta^{2} S}{\delta \a^{\ast} \delta \b} \biggr|_{\c}, \\ %%
   I_{\b_{\c}^{\ast}\a_{\c}} &= \frac{\delta^{2} S}{\delta \b^{\ast} \delta \a} \biggr|_{\c}, \\ %%
   I_{\b_{\c}^{\ast}\b_{\c}} &= \frac{\delta^{2} S}{\delta \b^{\ast} \delta \b} \biggr|_{\c},
\end{align}
with $I_{\a^{\ast}\a^{\ast}}$, $I_{\b^{\ast}\b^{\ast}}$, $\dots$ being functionals of $\a_{\c}^{\ast}$, $\b_{\c}^{\ast}$, $\a_{\c}$, and $\b_{\c}$.

In the followings, we decompose $\a_{\c}$ and $\b_{\c}$ into real and imaginary parts whose components are denoted by the subscripts $1$ and $2$, respectively, and denote $\varphi=\bigl(\a_{\c1} \, \a_{\c2} \, \b_{\c1} \, \b_{\c2}\bigr)^{t}$ due to the approximation, $\varphi \approx \phi_{\c}$, shown just after Eqs.~\eqref{eq:varphi_definition_1} and \eqref{eq:varphi_definition_2}.
With the indices by $\a_{\c1}$, $\a_{\c2}$, $\b_{\c1}$, and $\b_{\c2}$, the matrix ${\cal M}_{\c}$ in Eq.~\eqref{eq:action_Taylor_expansion} is given by
\begin{align}
   {\cal M}_{\c}
=
\left(
\begin{array}{cccc}
 M_{\a_{\c1}\a_{\c1} }& M_{\a_{\c1}\a_{2\c}} & M_{\a_{\c1}\b_{\c1}} & M_{\a_{\c1}\b_{\c2}} \\
 M_{\a_{\c2}\a_{\c1}} & M_{\a_{\c2}\a_{\c2}} & M_{\a_{\c2}\b_{\c1}} & M_{\a_{\c2}\b_{\c2}} \\
 M_{\b_{\c1}\a_{\c1}} & M_{\b_{\c1}\a_{\c2}} & M_{\b_{\c1}\b_{\c1}} & M_{\b_{\c1}\b_{\c2}} \\
 M_{\b_{\c2}\a_{\c1}} & M_{\b_{\c2}\a_{\c2}} & M_{\b_{\c2}\b_{\c1}} & M_{\b_{\c2}\b_{\c2}}
\end{array}
\right),
\end{align}
with the matrix components
\begin{align}
   M_{\a_{\c1}\a_{\c1}} &=
                                    I_{\a_{\c}^{\ast}\a_{\c}^{\ast}} + I_{\a_{\c}\a_{\c}} + I_{\a_{\c}^{\ast}\a_{\c}},
\\ %%
   M_{\a_{\c2}\a_{\c2}} &=
                                  - I_{\a_{\c}^{\ast}\a_{\c}^{\ast}} - I_{\a_{\c}\a_{\c}} + I_{\a_{\c}^{\ast}\a_{\c}},
\\ %%
   M_{\a_{\c1}\a_{\c2}} &= - iI_{\a_{\c}^{\ast}\a_{\c}^{\ast}} + iI_{\a_{\c}\a_{\c}},
\\ %%
    M_{\b_{\c1}\b_{\c1}} &=
                                     I_{\b_{\c}^{\ast}\b_{\c}^{\ast}} + I_{\b_{\c}\b_{\c}} + I_{\b_{\c}^{\ast}\b_{\c}},
\\ %%
   M_{\b_{\c2}\b_{\c2}} &=
                                  - I_{\b_{\c}^{\ast}\b_{\c}^{\ast}} - I_{\b_{\c}\b_{\c}} + I_{\b_{\c}^{\ast}\b_{\c}},
\\ %%
   M_{\b_{\c1}\b_{\c2}} &= - iI_{\b_{\c}^{\ast}\b_{\c}^{\ast}} + iI_{\b_{\c}\b_{\c}},
\\ %%
   M_{\a_{\c1}\b_{\c1}} &= \frac{1}{2} \bigl( I_{\a_{\c}^{\ast}\b_{\c}^{\ast}} + I_{\a_{\c}\b_{\c}} + I_{\a_{\c}^{\ast}\b_{\c}} + I_{\b_{\c}^{\ast}\a_{\c}} \bigr),
\\ %%
   M_{\a_{\c2}\b_{\c2}} &= \frac{1}{2} \bigl( - I_{\a_{\c}^{\ast}\b_{\c}^{\ast}} - I_{\a_{\c}\b_{\c}} + I_{\a_{\c}^{\ast}\b_{\c}} + I_{\b_{\c}^{\ast}\a_{\c}} \bigr),
\\ %%
   M_{\a_{\c1}\b_{\c2}} &= \frac{1}{2} \bigl( - iI_{\a_{\c}^{\ast}\b_{\c}^{\ast}} + iI_{\a_{\c}\b_{\c}} + iI_{\a_{\c}^{\ast}\b_{\c}} - iI_{\b_{\c}^{\ast}\a_{\c}} \bigr),
\\ %%
   M_{\a_{\c2}\b_{\c1}} &= \frac{1}{2} \bigl( - iI_{\a_{\c}^{\ast}\b_{\c}^{\ast}} + iI_{\a_{\c}\b_{\c}} - iI_{\a_{\c}^{\ast}\b_{\c}} + iI_{\b_{\c}^{\ast}\a_{\c}} \bigr).
\end{align}
Here $\tilde{\varphi}=\phi-\phi_{\c}$ and $\tilde{\varphi}^{\ast}=\phi^{\ast}-\phi_{\c}^{\ast}$ represent the small deviations of $\phi$ and $\phi^{\ast}$ from $\phi_{\c}$ and $\phi_{\c}^{\ast}$, respectively.
As the approximation in the last line in Eq.~\eqref{eq:action_Taylor_expansion}, we neglect the higher-order terms of $\tilde{\varphi}$ and $\tilde{\varphi}^{\ast}$.

Performing the path integral for $\tilde{\varphi}$ and $\tilde{\varphi}^{\ast}$ with the above approximation, we find that the generating functional~\eqref{eq:S_definition} is expressed by
\begin{widetext}
\begin{align}
   Z[j^{\ast},j]
&\approx
   {\cal N}
   \int {\cal D} \tilde{\varphi}^{\ast} {\cal D} \tilde{\varphi} \,
   \exp
   \Biggl(
       - S[\phi_{\c}^{\ast},\phi_{\c}]
      + \phi_{\c}^{\ast}j + \phi_{\c} j^{\ast}
       - \frac{1}{2}
         \frac{\delta^{2}S[\phi^{\ast},\phi]}{\delta \phi^{\ast 2}}
         \biggr|_{\c}
         \tilde{\varphi}^{\ast 2}
       - \frac{1}{2}
         \frac{\delta^{2}S[\phi^{\ast},\phi]}{\delta \phi^{2}}
         \biggr|_{\c}
         \tilde{\varphi}^{2}
       - \frac{\delta^{2}S[\phi^{\ast},\phi]}{\delta \phi^{\ast} \delta \phi}
         \biggr|_{\c}
         \tilde{\varphi}^{\ast}
         \tilde{\varphi}
   \Biggr)
\nonumber \\[0em] %%
&=
   {\cal N}
   \exp
   \Bigl(
       - S[\phi_{\c}^{\ast},\phi_{\c}]
      + \phi_{\c}^{\ast}j + \phi_{\c} j^{\ast}
   \Bigr)
   \int
   {\cal D} \tilde{\varphi}
   {\cal D} \tilde{\varphi}^{\ast}
   \exp
   \bigl(
       - \tilde{\varphi}^{t} {\cal M}_{\c}[\phi_{\c}^{\ast},\phi_{\c}] \tilde{\varphi}
   \bigr)
\nonumber \\[0em] %%
&=
   {\cal N}
   \exp
   \biggl(
       - S[\phi_{\c}^{\ast},\phi_{\c}]
      + \phi_{\c}^{\ast}j + \phi_{\c} j^{\ast}
       - \frac{1}{2} \Tr \ln {\cal M}_{\c}[\phi_{\c}^{\ast},\phi_{\c}]
   \biggr),
\end{align}
\end{widetext}
where the formula $\Dete \, A=\exp\bigl(\Tr \ln A\bigr)$ is used for the functional matrix $A$ with infinite dimensions over time and $\displaystyle \Tr \, F = \int_{0}^{t} \dr \tau \, F(\tau,\tau)$ is the trace for the functional matrix
 $F(\cdot,\cdot)$.
We explain the details of the trace calculation in Appendix~\ref{sec:trace_Matsubara_sum}.

With the above setups, the generating functional $W[j^{\ast},j] = \ln Z[j^{\ast},j]$ for the connected diagrams is given approximatetly by
\begin{align}
   W[j,j^{\ast}]
 =
 - S[\phi_{\c}^{\ast},\phi_{\c}]
+ \phi_{\c}^{\ast}j + \phi_{\c} j^{\ast}
 - \frac{1}{2} \Tr \ln {\cal M}_{\c}[\phi_{\c}^{\ast},\phi_{\c}].
\end{align}
Noting that $\phi_{\c}$ and $\phi_{\c}^{\ast}$ are the functionals of $j$ and $j^{\ast}$ through the equations of motion \eqref{eq:phi_EOM1} and \eqref{eq:phi_EOM2},
we express the effective action~\eqref{eq:effective_action_definition} by
\begin{align}
   \Gamma[\varphi^{\ast},\varphi]
&\approx
   S[\varphi^{\ast},\varphi]
+ \frac{1}{2} \Tr \ln {\cal M}_{\c}[\phi_{\c}^{\ast},\phi_{\c}]
\nonumber \\ %%
&=
   \int_{0}^{t} \dr \tau \, {\cal L}_{\eff}[\phi_{\c}^{\ast},\phi_{\c}].
\label{eq:effective_action_result}
\end{align}
In the last line in Eq.~\eqref{eq:effective_action_result}, we define the effective Lagrangian by
\begin{align}
   {\cal L}_{\eff}[\phi_{\c}^{\ast},\phi_{\c}]
&=
   \a_{\c}^{\ast} \partial_{\tau} \a_{\c}
   %\nonumber \\ & %%
+ \b_{\c}^{\ast} \partial_{\tau} \b_{\c}
   \nonumber \\ & %%
+ \biggl( \a_{\c}^{\ast} - \frac{1}{n}\b_{\c}^{\ast} \biggr)
   \bigl( \lambda\a_{\c}^{n} - \mu\b_{\c} \bigr)
   \nonumber \\ & %%
 - \biggl( \a_{\c}^{\ast} - \frac{1}{n}\b_{\c}^{\ast} \biggr)^{2}
   \nu \a^{l} \b^{m}
   \nonumber \\ & %%
+ \frac{1}{4} \sum_{k} \int_{0}^{m_{k}} \dr x \, \coth\biggl(\frac{t}{2}x\biggr)
   \nonumber \\ & \hspace{0em} %%s
+ \Biggl(
         \a_{\c0}^{\ast} \bigl( \a_{\c0} - \bar{n}_{\a0} \bigr)
         %\nonumber \\ & %%
      + \b_{\c0}^{\ast} \bigl( \b_{\c0} - \bar{n}_{\b0} \bigr)
         \nonumber \\ & %%
      + \frac{1}{2} \sum_{k} \ln \biggl( 1 + \frac{1}{2} \coth\biggl(\frac{t}{2}m_{k}\biggr) \biggr)
   \Biggr)
   \delta(\tau).
\end{align}
with $\a_{\c0}$ and $\b_{\c0}$ being $\a_{\c}(0)$ and $\b_{\c}(0)$ at $\tau=0$, respectively.
Here the eigenvalues of ${\cal M}_{\c}$ are given by $\partial_{\tau}+\delta(\tau)$, $\partial_{\tau}+\delta(\tau)$, $\partial_{\tau}+\delta(\tau)+m_{+}(\tau)$, $\partial_{\tau}+\delta(\tau)+m_{-}(\tau)$ with some functions $m_{\pm}(\tau)$, and the constant terms are neglected.
Considering the long-time scale ($t\rightarrow\infty$), we obtain the effective Lagrangian expressed by
\begin{align}
   {\cal L}_{\eff}[\phi_{\c}^{\ast},\phi_{\c}]
&\approx
   \a_{\c}^{\ast} \partial_{\tau} \a_{\c}
+ \b_{\c}^{\ast} \partial_{\tau} \b_{\c}
   \nonumber \\ & %%
+ \biggl( \a_{\c}^{\ast} - \frac{1}{n}\b_{\c}^{\ast} \biggr)
   \bigl( \lambda\a_{\c}^{n} - \mu\b_{\c} \bigr)
   \nonumber \\ & %%
 - \biggl( \a_{\c}^{\ast} - \frac{1}{n}\b_{\c}^{\ast} \biggr)^{2}
   \nu \a^{l} \b^{m}
   \nonumber \\ & \hspace{0em} %%
 - \nu
   \biggl( \a_{\c}^{\ast} - \frac{1}{n} \b_{\c}^{\ast} \biggr)
   \biggl( l \a_{\c}^{l-1}\b_{\c}^{m} - \frac{m}{n}\a_{\c}^{l}\b_{\c}^{m-1} \biggr)
   \nonumber \\ & %%
+ \a_{\c0}^{\ast} \bigl( \a_{\c0} - \bar{n}_{\a0} \bigr) \delta(\tau)
   %\nonumber \\ & %%
+ \b_{\c0}^{\ast} \bigl( \b_{\c0} - \bar{n}_{\b0} \bigr) \delta(\tau),
\label{eq:effective_Lagrangian_infinity}
\end{align}
where in the approximation we neglect the terms including no $\a_{\c}^{\ast}$ or $\b_{\c}^{\ast}$ because such terms are irrelevant to the equations of motion for $\a_{\c}$ and $\b_{\c}$.
Finally, we obtain the Euler-Lagrange equations from Eq.~\eqref{eq:effective_Lagrangian_infinity}, i.e., the equations of motion \eqref{eq:nAB2_model_WKB_A} and \eqref{eq:nAB2_model_WKB_B}.

\subsection{Calculation of trace and Matsubara sum} \label{sec:trace_Matsubara_sum}

In Eq.~\eqref{eq:effective_action_result}, we calculate the trace in $\Tr \ln {\cal M}_{\c}$ as a Matsubara sum in the following way:
\begin{align}
&
   \Tr \ln {\cal M}_{\c}[\phi_{\c}^{\ast},\phi_{\c}]
\nonumber \\ %%
&=
   \tr \int_{0}^{t} \dr \tau \, \ln {\cal M}_{\c}[\phi_{\c}^{\ast},\phi_{\c}](\tau,\tau)
\nonumber \\[0em] %%
&=
   \int_{0}^{t} \dr \tau \, \tr \ln {\cal M}_{\c}[\phi_{\c}^{\ast},\phi_{\c}](\tau,\tau)
\nonumber \\[0em] %%
&=
   \int_{0}^{t} \dr \tau \,
   \langle\tau|
   \ln
   \bigl(
         \partial_{\tau}
      + m_{+}
      + \delta(\tau)
   \bigr)
   |\tau\rangle
   \nonumber \\ & %%
+ \int_{0}^{t} \dr \tau \,
   \langle\tau|
   \ln
   \bigl(
         \partial_{\tau}
      + m_{-}
      + \delta(\tau)
   \bigr)
   |\tau\rangle
\nonumber \\[0em] %%
&=
   \sum_{\alpha=\pm}
   \int_{0}^{t} \dr \tau \,
   \langle\tau|
   \ln
   \Biggl(
         \bigl( \partial_{\tau} + m_{\alpha} \bigr)
         \biggl( 1+ \frac{\delta(\tau)}{\partial_{\tau} + m_{\alpha}} \biggr)
   \Biggr)
   |\tau\rangle
\nonumber \\[0em] %%
&=
   \sum_{\alpha=\pm}
   \int_{0}^{t} \dr \tau \,
   \langle\tau|
   \ln \bigl( \partial_{\tau} + m_{\alpha} \bigr)
   |\tau\rangle
   \nonumber \\ & %%
+ \sum_{\alpha=\pm}
   \int_{0}^{t} \dr \tau \,
   \langle\tau|
   \ln \biggl( 1+ \frac{\delta(\tau)}{\partial_{\tau} + m_{\alpha}} \biggr)
   |\tau\rangle
\nonumber \\[0em] %%
&=
   \sum_{\alpha=\pm}
   \int_{0}^{t} \dr \tau \,
   \frac{1}{2}
   \int_{0}^{m_{\alpha}} \dr x \,
   \coth\biggl(\frac{t}{2}x\biggr)
   \nonumber \\ & %%
+ \sum_{\alpha=\pm}
   \ln
   \Biggl(
         1
      + \frac{1}{2} \coth\biggl(\frac{t}{2}m_{\alpha}(0)\biggr)
   \Biggr).
\label{eq:TrlnM_formula}
\end{align}
In the last line in Eq.~\eqref{eq:TrlnM_formula}, we calculate the first term as
\begin{align}
&
   \int_{0}^{t} \dr \tau \,
   \langle\tau|
   \ln \bigl( \partial_{\tau} + m_{\alpha} \bigr)
   |\tau\rangle
\nonumber \\ %%
&=
   \int_{0}^{t} \dr \tau \,
   \langle\tau|
   \ln \bigl( \partial_{\tau} + m_{\alpha} \bigr)
   \frac{1}{t} \sum_{n} |\omega_{n}\rangle \langle\omega_{n}
   |\tau\rangle
\nonumber \\[0em] %%
&=
   \frac{1}{t}
   \sum_{n}
   \int_{0}^{t} \dr \tau \,
   \langle\tau|\omega_{n}\rangle
   \ln \bigl( \partial_{\tau} + m_{\alpha} \bigr)
   \langle\omega_{n}|\tau\rangle
\nonumber \\[0em] %%
&=
   \frac{1}{t}
   \sum_{n}
   \int_{0}^{t} \dr \tau \,
   e^{-i\omega_{n}\tau}
   \ln \bigl( \partial_{\tau} + m_{\alpha} \bigr)
   e^{i\omega_{n}\tau}
\nonumber \\[0em] %%
&=
   \frac{1}{t}
   \int_{0}^{t} \dr \tau \,
   \frac{1}{2}
   \sum_{n}
   \ln \bigl( \omega_{n}^{2} + m_{\alpha}^{2} \bigr)
\nonumber \\[0em] %%
&=
   \frac{1}{t}
   \int_{0}^{t} \dr \tau \,
   \frac{1}{2}
   \sum_{n}
   \int_{0}^{m_{\alpha}^{2}} \dr x^{2} \, \frac{1}{\omega_{n}^{2}+x^{2}}
\nonumber \\[0em] %%
&=
   \frac{1}{t}
   \int_{0}^{t} \dr \tau \,
   \frac{1}{2}
   \int_{0}^{m_{\alpha}^{2}} \dr x^{2}
   \sum_{n}
   \frac{1}{\omega_{n}^{2}+x^{2}}
\nonumber \\[0em] %%
&=
   \int_{0}^{t} \dr \tau \,
   \frac{1}{2}
   \int_{0}^{m_{\alpha}} \dr x \,
   \coth\biggl(\frac{t}{2}x\biggr),
\end{align}
noting that $m_{\alpha}=m_{\alpha}(\tau)$ with $\alpha=\pm$ is the function dependent on $\tau$ through $\phi^{\ast}(\tau)$ and $\phi(\tau)$.
In the first line of the above equations, we use the condition for complete system, i.e.,
$(1/t) \sum_{n} |\omega_{n}\rangle \langle\omega_{n}|=1$.\footnote{In quantum field theory, the conditions for the complete system for plane waves with position $x$ and momentum $p$ are given by
\begin{align}
   \int \dr x \, | x \rangle \langle x | = 1, \quad %%
   \int \frac{\dr p}{2\pi} \, | p \rangle \langle p | = 1,
\end{align}
and the inner product is given by $\langle x | p \rangle = e^{ipx}$ and $\langle p | x \rangle = e^{-ipx}$.
In the finite temperature field theory, analogously, we consider $\tau$ and $\omega_{n}$ instead of $x$ and $p$, respectively,  and impose the periodic boundary condition for $\tau$ in the interval $\tau \in [0,\beta]$, and introduce the discretization $\omega_{n}=2\pi n/\beta$ ($n \in \mathbb{Z}$)~\cite{Bellac:2011kqa,Kapusta:2023eix}.
Considering the correspondence between $| p \rangle$ and $| \omega_{n} \rangle$, we find that the condition for the complete system is discretized to be
\begin{align}
&
   \int \frac{\dr p}{2\pi} \, | p \rangle \langle p |
\nonumber \\ %%
&\Rightarrow
    \sum_{n} \frac{\Delta p}{2\pi} | p \rangle \langle p |
= \sum_{n} \frac{2\pi}{2\pi \beta} | \omega_{n} \rangle \langle \omega_{n} |
= \frac{1}{\beta} \sum_{n} | \omega_{n} \rangle \langle \omega_{n} |
= 1,
\end{align}
with $\Delta p=2\pi/\beta$, where we have replaced $|p\rangle$ by $|\omega_{n}\rangle$ in the middle of the calculation.
Regarding $\beta$ (inverse temperature) as $t$ (time) and imposing the periodic boundary condition for the interval $\tau \in [0,t]$, we obtain the condition for the complete system:
$(1/t) \sum_{n} |\omega_{n}\rangle \langle\omega_{n}|=1$
for $\omega_{n}=2\pi n / t$ ($n \in \mathbb{Z}$).}
Here $\omega_{n}=2\pi n / t$ ($n \in \mathbb{Z}$) is the Matsubara frequency,
and $\langle\tau|\omega_{n}\rangle=e^{-i\omega_{n}\tau}$ and $\langle\omega_{n}|\tau\rangle=e^{i\omega_{n}\tau}$ are used for ``plane waves".
We note that there is no need to determine the lower limit of the interval of the integral variable $x$, because the divergence of the Matsubara sum for $\ln(\cdots)$ can be subtracted appropriately without affecting the final result.
We calculate the Matsubara sum over $n$ by
\begin{align}
   \sum_{n}
   \frac{1}{\omega_{n}^{2}+x^{2}}
&=
   \sum_{n}
   \frac{1}{-(i\omega_{n})^{2}+x^{2}}
\nonumber \\[0em] %%
&=
 - \frac{t}{2\pi i} \int_{C_{f}} \dr z
   \sum_{C_{f}}
   \frac{1}{-z^{2}+x^{2}}
   \frac{1}{2} \coth\biggl(\frac{t}{2}z\biggr)
\nonumber \\[0em] %%
&=
   \frac{t}{2\pi i} \int_{C_{f}} \dr z
   \sum_{C_{f}}
   \frac{1}{z^{2}-x^{2}}
   \frac{1}{2} \coth\biggl(\frac{t}{2}z\biggr)
\nonumber \\[0em] %%
&=
   t
   \frac{1}{x} \frac{1}{2} \coth\biggl(\frac{t}{2}x\biggr),
\label{eq:Matsubara_sum_log_formula}
\end{align}
where we use the formula for the complex function $f(z)$ with a complex variable $z$,
\begin{align}
   \sum_{n} f(i\omega_{n})
=
 - \frac{t}{2\pi i}
   \int_{C_{f}} \dr z \, f(z) \frac{1}{2} \coth\biggl(\frac{t}{2}z\biggr),
\label{eq:Matsubara_sum_formula}
\end{align}
with $C_{f}$ being the sum of the anticlockwise paths around the poles of $f(z)$.
In Eq.~\eqref{eq:Matsubara_sum_formula}, we assume that the integrand becomes sufficiently zero at infinity ($|z|\rightarrow\infty$) and the contribution to the integral at the infinity can be neglected.
In Eq.~\eqref{eq:TrlnM_formula}, we calculate the second term as
\begin{align}
&
   \sum_{\alpha=\pm}
   \int_{0}^{t} \dr \tau \,
   \langle\tau|
   \ln \biggl( 1+ \frac{\delta(\tau)}{\partial_{\tau} + m_{\alpha}} \biggr)
   |\tau\rangle
\nonumber \\ %%
&=
   \sum_{\alpha=\pm}
   \int_{0}^{t} \dr \tau \,
   \langle\tau|
   \sum_{k\ge1}
   \frac{(-1)^{k+1}}{k} \biggl( \frac{\delta(\tau)}{\partial_{\tau} + m_{\alpha}} \biggr)^{k}
   |\tau\rangle
\nonumber \\[0em] %%
&=
   \sum_{\alpha=\pm}
   \sum_{k\ge1}
   \frac{(-1)^{k+1}}{k}
   \int_{0}^{t} \dr \tau \,
   \langle\tau|
   \biggl( \frac{\delta(\tau)}{\partial_{\tau} + m_{\alpha}} \biggr)^{k}
   |\tau\rangle
\nonumber \\[0em] %%
&=
   \sum_{\alpha=\pm}
   \ln
   \biggl(
         1
      + \frac{1}{t} \sum_{n} \frac{1}{i\omega_{n}+m_{\alpha}(0)}
   \biggr)
\nonumber \\[0em] %%
&=
   \sum_{\alpha=\pm}
   \ln
   \Biggl(
         1
      + \frac{1}{t}
         \frac{t}{2} \coth\biggl(\frac{t}{2}m_{\alpha}(0)\biggr)
   \Biggr)
\nonumber \\[0em] %%
&=
   \sum_{\alpha=\pm}
   \ln
   \Biggl(
         1
      + \frac{1}{2}
         \coth\biggl(\frac{t}{2}m_{\alpha}(0)\biggr)
   \Biggr),
\end{align}
where $m_{\alpha}(0)$ represents the value of $m_{\alpha}(\tau)$ at $\tau=0.$

\bibliography{reference}

%apsrev4-2.bst 2019-01-14 (MD) hand-edited version of apsrev4-1.bst
%Control: key (0)
%Control: author (72) initials jnrlst
%Control: editor formatted (1) identically to author
%Control: production of article title (-1) disabled
%Control: page (0) single
%Control: year (1) truncated
%Control: production of eprint (0) enabled
\begin{thebibliography}{51}%
\makeatletter
\providecommand \@ifxundefined [1]{%
 \@ifx{#1\undefined}
}%
\providecommand \@ifnum [1]{%
 \ifnum #1\expandafter \@firstoftwo
 \else \expandafter \@secondoftwo
 \fi
}%
\providecommand \@ifx [1]{%
 \ifx #1\expandafter \@firstoftwo
 \else \expandafter \@secondoftwo
 \fi
}%
\providecommand \natexlab [1]{#1}%
\providecommand \enquote  [1]{``#1''}%
\providecommand \bibnamefont  [1]{#1}%
\providecommand \bibfnamefont [1]{#1}%
\providecommand \citenamefont [1]{#1}%
\providecommand \href@noop [0]{\@secondoftwo}%
\providecommand \href [0]{\begingroup \@sanitize@url \@href}%
\providecommand \@href[1]{\@@startlink{#1}\@@href}%
\providecommand \@@href[1]{\endgroup#1\@@endlink}%
\providecommand \@sanitize@url [0]{\catcode `\\12\catcode `\$12\catcode
  `\&12\catcode `\#12\catcode `\^12\catcode `\_12\catcode `\%12\relax}%
\providecommand \@@startlink[1]{}%
\providecommand \@@endlink[0]{}%
\providecommand \url  [0]{\begingroup\@sanitize@url \@url }%
\providecommand \@url [1]{\endgroup\@href {#1}{\urlprefix }}%
\providecommand \urlprefix  [0]{URL }%
\providecommand \Eprint [0]{\href }%
\providecommand \doibase [0]{https://doi.org/}%
\providecommand \selectlanguage [0]{\@gobble}%
\providecommand \bibinfo  [0]{\@secondoftwo}%
\providecommand \bibfield  [0]{\@secondoftwo}%
\providecommand \translation [1]{[#1]}%
\providecommand \BibitemOpen [0]{}%
\providecommand \bibitemStop [0]{}%
\providecommand \bibitemNoStop [0]{.\EOS\space}%
\providecommand \EOS [0]{\spacefactor3000\relax}%
\providecommand \BibitemShut  [1]{\csname bibitem#1\endcsname}%
\let\auto@bib@innerbib\@empty
%</preamble>
\bibitem [{\citenamefont {Martin}\ \emph {et~al.}(1973)\citenamefont {Martin},
  \citenamefont {Siggia},\ and\ \citenamefont {Rose}}]{PhysRevA.8.423}%
  \BibitemOpen
  \bibfield  {author} {\bibinfo {author} {\bibfnamefont {P.~C.}\ \bibnamefont
  {Martin}}, \bibinfo {author} {\bibfnamefont {E.~D.}\ \bibnamefont {Siggia}},\
  and\ \bibinfo {author} {\bibfnamefont {H.~A.}\ \bibnamefont {Rose}},\ }\href
  {https://doi.org/10.1103/PhysRevA.8.423} {\bibfield  {journal} {\bibinfo
  {journal} {Phys. Rev. A}\ }\textbf {\bibinfo {volume} {8}},\ \bibinfo {pages}
  {423} (\bibinfo {year} {1973})}\BibitemShut {NoStop}%
\bibitem [{\citenamefont {de~Dominicis}(1976)}]{dedominicis:jpa-00216466}%
  \BibitemOpen
  \bibfield  {author} {\bibinfo {author} {\bibfnamefont {C.}~\bibnamefont
  {de~Dominicis}},\ }\href {https://doi.org/10.1051/jphyscol:1976138}
  {\bibfield  {journal} {\bibinfo  {journal} {{Journal de Physique Colloques}}\
  }\textbf {\bibinfo {volume} {37}},\ \bibinfo {pages} {C1} (\bibinfo {year}
  {1976})}\BibitemShut {NoStop}%
\bibitem [{\citenamefont {Janssen}(1976)}]{Janssen1976}%
  \BibitemOpen
  \bibfield  {author} {\bibinfo {author} {\bibfnamefont {H.-K.}\ \bibnamefont
  {Janssen}},\ }\href {https://doi.org/10.1007/BF01316547} {\bibfield
  {journal} {\bibinfo  {journal} {Zeitschrift f{\"u}r Physik B Condensed
  Matter}\ }\textbf {\bibinfo {volume} {23}},\ \bibinfo {pages} {377} (\bibinfo
  {year} {1976})}\BibitemShut {NoStop}%
\bibitem [{\citenamefont {Peliti}(1985)}]{Peliti_1985}%
  \BibitemOpen
  \bibfield  {author} {\bibinfo {author} {\bibfnamefont {L.}~\bibnamefont
  {Peliti}},\ }\href {https://doi.org/10.1051/jphys:019850046090146900}
  {\bibfield  {journal} {\bibinfo  {journal} {{Journal de Physique}}\ }\textbf
  {\bibinfo {volume} {46}},\ \bibinfo {pages} {1469} (\bibinfo {year}
  {1985})}\BibitemShut {NoStop}%
\bibitem [{\citenamefont {Doi}(1976{\natexlab{a}})}]{Doi1_1976}%
  \BibitemOpen
  \bibfield  {author} {\bibinfo {author} {\bibfnamefont {M.}~\bibnamefont
  {Doi}},\ }\href {https://doi.org/10.1088/0305-4470/9/9/008} {\bibfield
  {journal} {\bibinfo  {journal} {Journal of Physics A: Mathematical and
  General}\ }\textbf {\bibinfo {volume} {9}},\ \bibinfo {pages} {1465}
  (\bibinfo {year} {1976}{\natexlab{a}})}\BibitemShut {NoStop}%
\bibitem [{\citenamefont {Doi}(1976{\natexlab{b}})}]{Doi2_1976}%
  \BibitemOpen
  \bibfield  {author} {\bibinfo {author} {\bibfnamefont {M.}~\bibnamefont
  {Doi}},\ }\href {https://doi.org/10.1088/0305-4470/9/9/009} {\bibfield
  {journal} {\bibinfo  {journal} {Journal of Physics A: Mathematical and
  General}\ }\textbf {\bibinfo {volume} {9}},\ \bibinfo {pages} {1479}
  (\bibinfo {year} {1976}{\natexlab{b}})}\BibitemShut {NoStop}%
\bibitem [{\citenamefont {Grassberger}(1982)}]{Grassberger1982}%
  \BibitemOpen
  \bibfield  {author} {\bibinfo {author} {\bibfnamefont {P.}~\bibnamefont
  {Grassberger}},\ }\href {https://doi.org/10.1007/BF01313803} {\bibfield
  {journal} {\bibinfo  {journal} {Zeitschrift f{\"u}r Physik B Condensed
  Matter}\ }\textbf {\bibinfo {volume} {47}},\ \bibinfo {pages} {365} (\bibinfo
  {year} {1982})}\BibitemShut {NoStop}%
\bibitem [{\citenamefont {Droz}\ and\ \citenamefont
  {McKane}(1994)}]{Droz_1994}%
  \BibitemOpen
  \bibfield  {author} {\bibinfo {author} {\bibfnamefont {M.}~\bibnamefont
  {Droz}}\ and\ \bibinfo {author} {\bibfnamefont {A.}~\bibnamefont {McKane}},\
  }\href {https://doi.org/10.1088/0305-4470/27/13/002} {\bibfield  {journal}
  {\bibinfo  {journal} {Journal of Physics A: Mathematical and General}\
  }\textbf {\bibinfo {volume} {27}},\ \bibinfo {pages} {L467} (\bibinfo {year}
  {1994})}\BibitemShut {NoStop}%
\bibitem [{\citenamefont {Lee}\ and\ \citenamefont {Cardy}(1995)}]{Lee1995}%
  \BibitemOpen
  \bibfield  {author} {\bibinfo {author} {\bibfnamefont {B.~P.}\ \bibnamefont
  {Lee}}\ and\ \bibinfo {author} {\bibfnamefont {J.}~\bibnamefont {Cardy}},\
  }\href {https://doi.org/10.1007/BF02179861} {\bibfield  {journal} {\bibinfo
  {journal} {Journal of Statistical Physics}\ }\textbf {\bibinfo {volume}
  {80}},\ \bibinfo {pages} {971} (\bibinfo {year} {1995})}\BibitemShut
  {NoStop}%
\bibitem [{\citenamefont {Cardy}\ and\ \citenamefont
  {T\"auber}(1996)}]{Cardy1996}%
  \BibitemOpen
  \bibfield  {author} {\bibinfo {author} {\bibfnamefont {J.}~\bibnamefont
  {Cardy}}\ and\ \bibinfo {author} {\bibfnamefont {U.~C.}\ \bibnamefont
  {T\"auber}},\ }\href {https://doi.org/10.1103/PhysRevLett.77.4780} {\bibfield
   {journal} {\bibinfo  {journal} {Phys. Rev. Lett.}\ }\textbf {\bibinfo
  {volume} {77}},\ \bibinfo {pages} {4780} (\bibinfo {year}
  {1996})}\BibitemShut {NoStop}%
\bibitem [{\citenamefont {{Cardy}}(1996)}]{cardy1996renormalisation}%
  \BibitemOpen
  \bibfield  {author} {\bibinfo {author} {\bibfnamefont {J.}~\bibnamefont
  {{Cardy}}},\ }\href@noop {} {\bibinfo {title} {Renormalisation group approach
  to reaction-diffusion problems}} (\bibinfo {year} {1996}),\ \Eprint
  {https://arxiv.org/abs/cond-mat/9607163} {arXiv:cond-mat/9607163 [cond-mat]}
  \BibitemShut {NoStop}%
\bibitem [{\citenamefont {Bettelheim}\ \emph {et~al.}(2001)\citenamefont
  {Bettelheim}, \citenamefont {Agam},\ and\ \citenamefont
  {Shnerb}}]{Bettelheim2001}%
  \BibitemOpen
  \bibfield  {author} {\bibinfo {author} {\bibfnamefont {E.}~\bibnamefont
  {Bettelheim}}, \bibinfo {author} {\bibfnamefont {O.}~\bibnamefont {Agam}},\
  and\ \bibinfo {author} {\bibfnamefont {N.~M.}\ \bibnamefont {Shnerb}},\
  }\href {https://doi.org/https://doi.org/10.1016/S1386-9477(00)00268-X}
  {\bibfield  {journal} {\bibinfo  {journal} {Physica E: Low-dimensional
  Systems and Nanostructures}\ }\textbf {\bibinfo {volume} {9}},\ \bibinfo
  {pages} {600} (\bibinfo {year} {2001})},\ \bibinfo {note} {proceedings of an
  International Workshop and Seminar on the Dynamics of Complex
  Systems}\BibitemShut {NoStop}%
\bibitem [{\citenamefont {Pastor-Satorras}\ and\ \citenamefont
  {Sol\'e}(2001)}]{Pastor-Satorras2001}%
  \BibitemOpen
  \bibfield  {author} {\bibinfo {author} {\bibfnamefont {R.}~\bibnamefont
  {Pastor-Satorras}}\ and\ \bibinfo {author} {\bibfnamefont {R.~V.}\
  \bibnamefont {Sol\'e}},\ }\href {https://doi.org/10.1103/PhysRevE.64.051909}
  {\bibfield  {journal} {\bibinfo  {journal} {Phys. Rev. E}\ }\textbf {\bibinfo
  {volume} {64}},\ \bibinfo {pages} {051909} (\bibinfo {year}
  {2001})}\BibitemShut {NoStop}%
\bibitem [{\citenamefont {Elgart}\ and\ \citenamefont
  {Kamenev}(2004)}]{Elgart2004}%
  \BibitemOpen
  \bibfield  {author} {\bibinfo {author} {\bibfnamefont {V.}~\bibnamefont
  {Elgart}}\ and\ \bibinfo {author} {\bibfnamefont {A.}~\bibnamefont
  {Kamenev}},\ }\href {https://doi.org/10.1103/PhysRevE.70.041106} {\bibfield
  {journal} {\bibinfo  {journal} {Phys. Rev. E}\ }\textbf {\bibinfo {volume}
  {70}},\ \bibinfo {pages} {041106} (\bibinfo {year} {2004})}\BibitemShut
  {NoStop}%
\bibitem [{\citenamefont {Elgart}\ and\ \citenamefont
  {Kamenev}(2006)}]{Elgart2006}%
  \BibitemOpen
  \bibfield  {author} {\bibinfo {author} {\bibfnamefont {V.}~\bibnamefont
  {Elgart}}\ and\ \bibinfo {author} {\bibfnamefont {A.}~\bibnamefont
  {Kamenev}},\ }\href {https://doi.org/10.1103/PhysRevE.74.041101} {\bibfield
  {journal} {\bibinfo  {journal} {Phys. Rev. E}\ }\textbf {\bibinfo {volume}
  {74}},\ \bibinfo {pages} {041101} (\bibinfo {year} {2006})}\BibitemShut
  {NoStop}%
\bibitem [{\citenamefont {Andreanov}\ \emph {et~al.}(2006)\citenamefont
  {Andreanov}, \citenamefont {Biroli}, \citenamefont {Bouchaud},\ and\
  \citenamefont {Lef\`evre}}]{Andreanov2006}%
  \BibitemOpen
  \bibfield  {author} {\bibinfo {author} {\bibfnamefont {A.}~\bibnamefont
  {Andreanov}}, \bibinfo {author} {\bibfnamefont {G.}~\bibnamefont {Biroli}},
  \bibinfo {author} {\bibfnamefont {J.-P.}\ \bibnamefont {Bouchaud}},\ and\
  \bibinfo {author} {\bibfnamefont {A.}~\bibnamefont {Lef\`evre}},\ }\href
  {https://doi.org/10.1103/PhysRevE.74.030101} {\bibfield  {journal} {\bibinfo
  {journal} {Phys. Rev. E}\ }\textbf {\bibinfo {volume} {74}},\ \bibinfo
  {pages} {030101} (\bibinfo {year} {2006})}\BibitemShut {NoStop}%
\bibitem [{\citenamefont {Mobilia}\ \emph {et~al.}(2007)\citenamefont
  {Mobilia}, \citenamefont {Georgiev},\ and\ \citenamefont
  {T{\"a}uber}}]{Mobilia2007}%
  \BibitemOpen
  \bibfield  {author} {\bibinfo {author} {\bibfnamefont {M.}~\bibnamefont
  {Mobilia}}, \bibinfo {author} {\bibfnamefont {I.~T.}\ \bibnamefont
  {Georgiev}},\ and\ \bibinfo {author} {\bibfnamefont {U.~C.}\ \bibnamefont
  {T{\"a}uber}},\ }\href {https://doi.org/10.1007/s10955-006-9146-3} {\bibfield
   {journal} {\bibinfo  {journal} {Journal of Statistical Physics}\ }\textbf
  {\bibinfo {volume} {128}},\ \bibinfo {pages} {447} (\bibinfo {year}
  {2007})}\BibitemShut {NoStop}%
\bibitem [{\citenamefont {Butler}\ and\ \citenamefont
  {Reynolds}(2009)}]{Butler2009}%
  \BibitemOpen
  \bibfield  {author} {\bibinfo {author} {\bibfnamefont {T.}~\bibnamefont
  {Butler}}\ and\ \bibinfo {author} {\bibfnamefont {D.}~\bibnamefont
  {Reynolds}},\ }\href {https://doi.org/10.1103/PhysRevE.79.032901} {\bibfield
  {journal} {\bibinfo  {journal} {Phys. Rev. E}\ }\textbf {\bibinfo {volume}
  {79}},\ \bibinfo {pages} {032901} (\bibinfo {year} {2009})}\BibitemShut
  {NoStop}%
\bibitem [{\citenamefont {T{\"a}uber}(2011)}]{Tauber_2011}%
  \BibitemOpen
  \bibfield  {author} {\bibinfo {author} {\bibfnamefont {U.~C.}\ \bibnamefont
  {T{\"a}uber}},\ }\href {https://doi.org/10.1088/1742-6596/319/1/012019}
  {\bibfield  {journal} {\bibinfo  {journal} {Journal of Physics: Conference
  Series}\ }\textbf {\bibinfo {volume} {319}},\ \bibinfo {pages} {012019}
  (\bibinfo {year} {2011})}\BibitemShut {NoStop}%
\bibitem [{\citenamefont {T{\"a}uber}(2012)}]{Tauber_2012}%
  \BibitemOpen
  \bibfield  {author} {\bibinfo {author} {\bibfnamefont {U.~C.}\ \bibnamefont
  {T{\"a}uber}},\ }\href {https://doi.org/10.1088/1751-8113/45/40/405002}
  {\bibfield  {journal} {\bibinfo  {journal} {Journal of Physics A:
  Mathematical and Theoretical}\ }\textbf {\bibinfo {volume} {45}},\ \bibinfo
  {pages} {405002} (\bibinfo {year} {2012})}\BibitemShut {NoStop}%
\bibitem [{\citenamefont {Oizumi}\ and\ \citenamefont
  {Takada}(2013)}]{Oizumi2013}%
  \BibitemOpen
  \bibfield  {author} {\bibinfo {author} {\bibfnamefont {R.}~\bibnamefont
  {Oizumi}}\ and\ \bibinfo {author} {\bibfnamefont {T.}~\bibnamefont
  {Takada}},\ }\href
  {https://doi.org/https://doi.org/10.1016/j.jtbi.2013.01.020} {\bibfield
  {journal} {\bibinfo  {journal} {Journal of Theoretical Biology}\ }\textbf
  {\bibinfo {volume} {323}},\ \bibinfo {pages} {76} (\bibinfo {year}
  {2013})}\BibitemShut {NoStop}%
\bibitem [{\citenamefont {Shih}\ and\ \citenamefont
  {Goldenfeld}(2014)}]{Shih2014}%
  \BibitemOpen
  \bibfield  {author} {\bibinfo {author} {\bibfnamefont {H.-Y.}\ \bibnamefont
  {Shih}}\ and\ \bibinfo {author} {\bibfnamefont {N.}~\bibnamefont
  {Goldenfeld}},\ }\href {https://doi.org/10.1103/PhysRevE.90.050702}
  {\bibfield  {journal} {\bibinfo  {journal} {Phys. Rev. E}\ }\textbf {\bibinfo
  {volume} {90}},\ \bibinfo {pages} {050702} (\bibinfo {year}
  {2014})}\BibitemShut {NoStop}%
\bibitem [{\citenamefont {Dickman}\ and\ \citenamefont
  {Vidigal}(2003)}]{Dickman2003}%
  \BibitemOpen
  \bibfield  {author} {\bibinfo {author} {\bibfnamefont {R.}~\bibnamefont
  {Dickman}}\ and\ \bibinfo {author} {\bibfnamefont {R.}~\bibnamefont
  {Vidigal}},\ }\href {https://doi.org/10.1590/S0103-97332003000100005}
  {\bibfield  {journal} {\bibinfo  {journal} {Braz. J. Phys.}\ }\textbf
  {\bibinfo {volume} {33}},\ \bibinfo {pages} {73} (\bibinfo {year}
  {2003})}\BibitemShut {NoStop}%
\bibitem [{\citenamefont {Janssen}\ and\ \citenamefont
  {T{\"a}uber}(2005)}]{Janssen2005}%
  \BibitemOpen
  \bibfield  {author} {\bibinfo {author} {\bibfnamefont {H.-K.}\ \bibnamefont
  {Janssen}}\ and\ \bibinfo {author} {\bibfnamefont {U.~C.}\ \bibnamefont
  {T{\"a}uber}},\ }\href {https://doi.org/10.1016/j.aop.2004.09.011} {\bibfield
   {journal} {\bibinfo  {journal} {Annals of Physics}\ }\textbf {\bibinfo
  {volume} {315}},\ \bibinfo {pages} {147} (\bibinfo {year}
  {2005})}\BibitemShut {NoStop}%
\bibitem [{\citenamefont {T{\"a}uber}\ \emph {et~al.}(2005)\citenamefont
  {T{\"a}uber}, \citenamefont {Howard},\ and\ \citenamefont
  {Vollmayr-Lee}}]{Tauber_2005}%
  \BibitemOpen
  \bibfield  {author} {\bibinfo {author} {\bibfnamefont {U.~C.}\ \bibnamefont
  {T{\"a}uber}}, \bibinfo {author} {\bibfnamefont {M.}~\bibnamefont {Howard}},\
  and\ \bibinfo {author} {\bibfnamefont {B.~P.}\ \bibnamefont {Vollmayr-Lee}},\
  }\href {https://doi.org/10.1088/0305-4470/38/17/r01} {\bibfield  {journal}
  {\bibinfo  {journal} {Journal of Physics A: Mathematical and General}\
  }\textbf {\bibinfo {volume} {38}},\ \bibinfo {pages} {R79} (\bibinfo {year}
  {2005})}\BibitemShut {NoStop}%
\bibitem [{\citenamefont {Cardy}\ \emph {et~al.}(2008)\citenamefont {Cardy},
  \citenamefont {Cardy}, \citenamefont {Falkovich},\ and\ \citenamefont
  {Gawedzki}}]{cardy_cardy_falkovich_gawedzki_2008}%
  \BibitemOpen
  \bibfield  {author} {\bibinfo {author} {\bibfnamefont {J.}~\bibnamefont
  {Cardy}}, \bibinfo {author} {\bibfnamefont {J.}~\bibnamefont {Cardy}},
  \bibinfo {author} {\bibfnamefont {G.}~\bibnamefont {Falkovich}},\ and\
  \bibinfo {author} {\bibfnamefont {K.}~\bibnamefont {Gawedzki}},\ }\bibinfo
  {title} {John cardy. reaction-diffusion processes},\ in\ \href
  {https://doi.org/10.1017/CBO9780511812149.004} {\emph {\bibinfo {booktitle}
  {Non-equilibrium Statistical Mechanics and Turbulence}}},\ \bibinfo {series
  and number} {London Mathematical Society Lecture Note Series},\ \bibinfo
  {editor} {edited by\ \bibinfo {editor} {\bibfnamefont {S.}~\bibnamefont
  {Nazarenko}}\ and\ \bibinfo {editor} {\bibfnamefont {O.~V.}\ \bibnamefont
  {Zaboronski}}}\ (\bibinfo  {publisher} {Cambridge University Press},\
  \bibinfo {year} {2008})\ pp.\ \bibinfo {pages} {108--161}\BibitemShut
  {NoStop}%
\bibitem [{\citenamefont {T{\"a}uber}(2009)}]{Tauber2009}%
  \BibitemOpen
  \bibfield  {author} {\bibinfo {author} {\bibfnamefont {U.~C.}\ \bibnamefont
  {T{\"a}uber}},\ }\bibinfo {title} {Field theoretic methods},\ in\ \href
  {https://doi.org/10.1007/978-0-387-30440-3_200} {\emph {\bibinfo {booktitle}
  {Encyclopedia of Complexity and Systems Science}}},\ \bibinfo {editor}
  {edited by\ \bibinfo {editor} {\bibfnamefont {R.~A.}\ \bibnamefont
  {Meyers}}}\ (\bibinfo  {publisher} {Springer New York},\ \bibinfo {address}
  {New York, NY},\ \bibinfo {year} {2009})\ pp.\ \bibinfo {pages}
  {3360--3374}\BibitemShut {NoStop}%
\bibitem [{\citenamefont {T{\"a}uber}(2014)}]{Tauber_2014}%
  \BibitemOpen
  \bibfield  {author} {\bibinfo {author} {\bibfnamefont {U.~C.}\ \bibnamefont
  {T{\"a}uber}},\ }\href {https://doi.org/10.1017/CBO9781139046213} {\emph
  {\bibinfo {title} {Critical Dynamics: A Field Theory Approach to Equilibrium
  and Non-Equilibrium Scaling Behavior}}}\ (\bibinfo  {publisher} {Cambridge
  University Press},\ \bibinfo {year} {2014})\BibitemShut {NoStop}%
\bibitem [{\citenamefont {Weber}\ and\ \citenamefont
  {Frey}(2017)}]{Weber_2017}%
  \BibitemOpen
  \bibfield  {author} {\bibinfo {author} {\bibfnamefont {M.~F.}\ \bibnamefont
  {Weber}}\ and\ \bibinfo {author} {\bibfnamefont {E.}~\bibnamefont {Frey}},\
  }\href {https://doi.org/10.1088/1361-6633/aa5ae2} {\bibfield  {journal}
  {\bibinfo  {journal} {Reports on Progress in Physics}\ }\textbf {\bibinfo
  {volume} {80}},\ \bibinfo {pages} {046601} (\bibinfo {year}
  {2017})}\BibitemShut {NoStop}%
\bibitem [{\citenamefont {Bressloff}(2014)}]{Bressloff2014}%
  \BibitemOpen
  \bibfield  {author} {\bibinfo {author} {\bibfnamefont {P.~C.}\ \bibnamefont
  {Bressloff}},\ }\href
  {https://doi.org/https://doi.org/10.1007/978-3-319-08488-6} {\emph {\bibinfo
  {title} {Stochastic Processes in Cell Biology}}}\ (\bibinfo  {publisher}
  {Springer},\ \bibinfo {year} {2014})\BibitemShut {NoStop}%
\bibitem [{\citenamefont {Aparicio}\ and\ \citenamefont
  {Solari}(2001)}]{PhysRevLett.86.4183}%
  \BibitemOpen
  \bibfield  {author} {\bibinfo {author} {\bibfnamefont {J.~P.}\ \bibnamefont
  {Aparicio}}\ and\ \bibinfo {author} {\bibfnamefont {H.~G.}\ \bibnamefont
  {Solari}},\ }\href {https://doi.org/10.1103/PhysRevLett.86.4183} {\bibfield
  {journal} {\bibinfo  {journal} {Phys. Rev. Lett.}\ }\textbf {\bibinfo
  {volume} {86}},\ \bibinfo {pages} {4183} (\bibinfo {year}
  {2001})}\BibitemShut {NoStop}%
\bibitem [{\citenamefont {Tom\'e}\ and\ \citenamefont
  {de~Oliveira}(2009)}]{PhysRevE.79.061128}%
  \BibitemOpen
  \bibfield  {author} {\bibinfo {author} {\bibfnamefont {T.}~\bibnamefont
  {Tom\'e}}\ and\ \bibinfo {author} {\bibfnamefont {M.~J.}\ \bibnamefont
  {de~Oliveira}},\ }\href {https://doi.org/10.1103/PhysRevE.79.061128}
  {\bibfield  {journal} {\bibinfo  {journal} {Phys. Rev. E}\ }\textbf {\bibinfo
  {volume} {79}},\ \bibinfo {pages} {061128} (\bibinfo {year}
  {2009})}\BibitemShut {NoStop}%
\bibitem [{\citenamefont {Ovaskainen}\ and\ \citenamefont
  {Meerson}(2010)}]{OVASKAINEN2010643}%
  \BibitemOpen
  \bibfield  {author} {\bibinfo {author} {\bibfnamefont {O.}~\bibnamefont
  {Ovaskainen}}\ and\ \bibinfo {author} {\bibfnamefont {B.}~\bibnamefont
  {Meerson}},\ }\href
  {https://doi.org/https://doi.org/10.1016/j.tree.2010.07.009} {\bibfield
  {journal} {\bibinfo  {journal} {Trends in Ecology \& Evolution}\ }\textbf
  {\bibinfo {volume} {25}},\ \bibinfo {pages} {643} (\bibinfo {year}
  {2010})}\BibitemShut {NoStop}%
\bibitem [{\citenamefont {Zhang}\ and\ \citenamefont
  {Wolynes}(2014)}]{Zhang2014}%
  \BibitemOpen
  \bibfield  {author} {\bibinfo {author} {\bibfnamefont {B.}~\bibnamefont
  {Zhang}}\ and\ \bibinfo {author} {\bibfnamefont {P.~G.}\ \bibnamefont
  {Wolynes}},\ }\href {https://doi.org/10.1073/pnas.1408561111} {\bibfield
  {journal} {\bibinfo  {journal} {Proceedings of the National Academy of
  Sciences}\ }\textbf {\bibinfo {volume} {111}},\ \bibinfo {pages} {10185}
  (\bibinfo {year} {2014})}\BibitemShut {NoStop}%
\bibitem [{\citenamefont {Stapmanns}\ \emph {et~al.}(2020)\citenamefont
  {Stapmanns}, \citenamefont {K\"uhn}, \citenamefont {Dahmen}, \citenamefont
  {Luu}, \citenamefont {Honerkamp},\ and\ \citenamefont
  {Helias}}]{PhysRevE.101.042124}%
  \BibitemOpen
  \bibfield  {author} {\bibinfo {author} {\bibfnamefont {J.}~\bibnamefont
  {Stapmanns}}, \bibinfo {author} {\bibfnamefont {T.}~\bibnamefont {K\"uhn}},
  \bibinfo {author} {\bibfnamefont {D.}~\bibnamefont {Dahmen}}, \bibinfo
  {author} {\bibfnamefont {T.}~\bibnamefont {Luu}}, \bibinfo {author}
  {\bibfnamefont {C.}~\bibnamefont {Honerkamp}},\ and\ \bibinfo {author}
  {\bibfnamefont {M.}~\bibnamefont {Helias}},\ }\href
  {https://doi.org/10.1103/PhysRevE.101.042124} {\bibfield  {journal} {\bibinfo
   {journal} {Phys. Rev. E}\ }\textbf {\bibinfo {volume} {101}},\ \bibinfo
  {pages} {042124} (\bibinfo {year} {2020})}\BibitemShut {NoStop}%
\bibitem [{\citenamefont {Stapmanns}\ \emph {et~al.}(2022)\citenamefont
  {Stapmanns}, \citenamefont {K\"uhn}, \citenamefont {Dahmen}, \citenamefont
  {Luu}, \citenamefont {Honerkamp},\ and\ \citenamefont
  {Helias}}]{PhysRevE.105.059901}%
  \BibitemOpen
  \bibfield  {author} {\bibinfo {author} {\bibfnamefont {J.}~\bibnamefont
  {Stapmanns}}, \bibinfo {author} {\bibfnamefont {T.}~\bibnamefont {K\"uhn}},
  \bibinfo {author} {\bibfnamefont {D.}~\bibnamefont {Dahmen}}, \bibinfo
  {author} {\bibfnamefont {T.}~\bibnamefont {Luu}}, \bibinfo {author}
  {\bibfnamefont {C.}~\bibnamefont {Honerkamp}},\ and\ \bibinfo {author}
  {\bibfnamefont {M.}~\bibnamefont {Helias}},\ }\href
  {https://doi.org/10.1103/PhysRevE.105.059901} {\bibfield  {journal} {\bibinfo
   {journal} {Phys. Rev. E}\ }\textbf {\bibinfo {volume} {105}},\ \bibinfo
  {pages} {059901} (\bibinfo {year} {2022})}\BibitemShut {NoStop}%
\bibitem [{\citenamefont {Yasui}\ \emph {et~al.}(2022)\citenamefont {Yasui},
  \citenamefont {Hatakeyama},\ and\ \citenamefont
  {Okuhara}}]{yasui2022criticality}%
  \BibitemOpen
  \bibfield  {author} {\bibinfo {author} {\bibfnamefont {S.}~\bibnamefont
  {Yasui}}, \bibinfo {author} {\bibfnamefont {Y.}~\bibnamefont {Hatakeyama}},\
  and\ \bibinfo {author} {\bibfnamefont {Y.}~\bibnamefont {Okuhara}},\
  }\href@noop {} {\bibinfo {title} {Criticality in stochastic sir model for
  infectious diseases}} (\bibinfo {year} {2022}),\ \Eprint
  {https://arxiv.org/abs/2202.05468} {arXiv:2202.05468 [q-bio.PE]} \BibitemShut
  {NoStop}%
\bibitem [{\citenamefont {Horsthemke}\ and\ \citenamefont
  {Lefever}(1984)}]{Horsthemke_Lefever_1998}%
  \BibitemOpen
  \bibfield  {author} {\bibinfo {author} {\bibfnamefont {W.}~\bibnamefont
  {Horsthemke}}\ and\ \bibinfo {author} {\bibfnamefont {R.}~\bibnamefont
  {Lefever}},\ }\href {https://doi.org/10.1007/3-540-36852-3} {\emph {\bibinfo
  {title} {Noise-Induced Transitions}}}\ (\bibinfo  {publisher}
  {Springer-Verlag Berlin Heidelberg},\ \bibinfo {year} {1984})\BibitemShut
  {NoStop}%
\bibitem [{\citenamefont {Garc{\'{\i}}a-Ojalvo}\ and\ \citenamefont
  {Sancho}(1999)}]{Garc_a_Ojalvo_1999}%
  \BibitemOpen
  \bibfield  {author} {\bibinfo {author} {\bibfnamefont {J.}~\bibnamefont
  {Garc{\'{\i}}a-Ojalvo}}\ and\ \bibinfo {author} {\bibfnamefont {J.~M.}\
  \bibnamefont {Sancho}},\ }\href {https://doi.org/10.1007/978-1-4612-1536-3}
  {\emph {\bibinfo {title} {Noise-Induced Phase Transitions}}}\ (\bibinfo
  {publisher} {Springer New York},\ \bibinfo {year} {1999})\BibitemShut
  {NoStop}%
\bibitem [{\citenamefont {Schenzle}\ and\ \citenamefont
  {Brand}(1979)}]{PhysRevA.20.1628}%
  \BibitemOpen
  \bibfield  {author} {\bibinfo {author} {\bibfnamefont {A.}~\bibnamefont
  {Schenzle}}\ and\ \bibinfo {author} {\bibfnamefont {H.}~\bibnamefont
  {Brand}},\ }\href {https://doi.org/10.1103/PhysRevA.20.1628} {\bibfield
  {journal} {\bibinfo  {journal} {Phys. Rev. A}\ }\textbf {\bibinfo {volume}
  {20}},\ \bibinfo {pages} {1628} (\bibinfo {year} {1979})}\BibitemShut
  {NoStop}%
\bibitem [{\citenamefont {Suzuki}\ \emph {et~al.}(1981)\citenamefont {Suzuki},
  \citenamefont {Kaneko},\ and\ \citenamefont {Sasagawa}}]{10.1143/PTP.65.828}%
  \BibitemOpen
  \bibfield  {author} {\bibinfo {author} {\bibfnamefont {M.}~\bibnamefont
  {Suzuki}}, \bibinfo {author} {\bibfnamefont {K.}~\bibnamefont {Kaneko}},\
  and\ \bibinfo {author} {\bibfnamefont {F.}~\bibnamefont {Sasagawa}},\ }\href
  {https://doi.org/10.1143/PTP.65.828} {\bibfield  {journal} {\bibinfo
  {journal} {Progress of Theoretical Physics}\ }\textbf {\bibinfo {volume}
  {65}},\ \bibinfo {pages} {828} (\bibinfo {year} {1981})}\BibitemShut
  {NoStop}%
\bibitem [{\citenamefont {Biancalani}\ \emph {et~al.}(2012)\citenamefont
  {Biancalani}, \citenamefont {Rogers},\ and\ \citenamefont
  {McKane}}]{PhysRevE.86.010106}%
  \BibitemOpen
  \bibfield  {author} {\bibinfo {author} {\bibfnamefont {T.}~\bibnamefont
  {Biancalani}}, \bibinfo {author} {\bibfnamefont {T.}~\bibnamefont {Rogers}},\
  and\ \bibinfo {author} {\bibfnamefont {A.~J.}\ \bibnamefont {McKane}},\
  }\href {https://doi.org/10.1103/PhysRevE.86.010106} {\bibfield  {journal}
  {\bibinfo  {journal} {Phys. Rev. E}\ }\textbf {\bibinfo {volume} {86}},\
  \bibinfo {pages} {010106} (\bibinfo {year} {2012})}\BibitemShut {NoStop}%
\bibitem [{\citenamefont {Ohkubo}\ \emph {et~al.}(2008)\citenamefont {Ohkubo},
  \citenamefont {Shnerb},\ and\ \citenamefont
  {A.~Kessler}}]{doi:10.1143/JPSJ.77.044002}%
  \BibitemOpen
  \bibfield  {author} {\bibinfo {author} {\bibfnamefont {J.}~\bibnamefont
  {Ohkubo}}, \bibinfo {author} {\bibfnamefont {N.}~\bibnamefont {Shnerb}},\
  and\ \bibinfo {author} {\bibfnamefont {D.}~\bibnamefont {A.~Kessler}},\
  }\href {https://doi.org/10.1143/JPSJ.77.044002} {\bibfield  {journal}
  {\bibinfo  {journal} {Journal of the Physical Society of Japan}\ }\textbf
  {\bibinfo {volume} {77}},\ \bibinfo {pages} {044002} (\bibinfo {year}
  {2008})}\BibitemShut {NoStop}%
\bibitem [{\citenamefont {Kashiwagi}\ \emph {et~al.}(2006)\citenamefont
  {Kashiwagi}, \citenamefont {Urabe}, \citenamefont {Kaneko},\ and\
  \citenamefont {Yomo}}]{10.1371/journal.pone.0000049}%
  \BibitemOpen
  \bibfield  {author} {\bibinfo {author} {\bibfnamefont {A.}~\bibnamefont
  {Kashiwagi}}, \bibinfo {author} {\bibfnamefont {I.}~\bibnamefont {Urabe}},
  \bibinfo {author} {\bibfnamefont {K.}~\bibnamefont {Kaneko}},\ and\ \bibinfo
  {author} {\bibfnamefont {T.}~\bibnamefont {Yomo}},\ }\href
  {https://doi.org/10.1371/journal.pone.0000049} {\bibfield  {journal}
  {\bibinfo  {journal} {PLOS ONE}\ }\textbf {\bibinfo {volume} {1}},\ \bibinfo
  {pages} {1} (\bibinfo {year} {2006})}\BibitemShut {NoStop}%
\bibitem [{\citenamefont {Onsager}\ and\ \citenamefont
  {Machlup}(1953)}]{PhysRev.91.1505}%
  \BibitemOpen
  \bibfield  {author} {\bibinfo {author} {\bibfnamefont {L.}~\bibnamefont
  {Onsager}}\ and\ \bibinfo {author} {\bibfnamefont {S.}~\bibnamefont
  {Machlup}},\ }\href {https://doi.org/10.1103/PhysRev.91.1505} {\bibfield
  {journal} {\bibinfo  {journal} {Phys. Rev.}\ }\textbf {\bibinfo {volume}
  {91}},\ \bibinfo {pages} {1505} (\bibinfo {year} {1953})}\BibitemShut
  {NoStop}%
\bibitem [{\citenamefont {Hongler}(1979)}]{Hongler1979}%
  \BibitemOpen
  \bibfield  {author} {\bibinfo {author} {\bibfnamefont {M.~O.}\ \bibnamefont
  {Hongler}},\ }\href {https://doi.org/10.5169/seals-115031} {\bibfield
  {journal} {\bibinfo  {journal} {Helvetica Physica Acta}\ }\textbf {\bibinfo
  {volume} {52}},\ \bibinfo {pages} {280} (\bibinfo {year} {1979})}\BibitemShut
  {NoStop}%
\bibitem [{\citenamefont {Toral}(2011)}]{Toral_2011}%
  \BibitemOpen
  \bibfield  {author} {\bibinfo {author} {\bibfnamefont {R.}~\bibnamefont
  {Toral}},\ }\href {https://doi.org/10.1063/1.3569493} {\bibfield  {journal}
  {\bibinfo  {journal} {AIP Conference Proceedings}\ }\textbf {\bibinfo
  {volume} {1332}},\ \bibinfo {pages} {145} (\bibinfo {year}
  {2011})}\BibitemShut {NoStop}%
\bibitem [{\citenamefont {Meerson}\ and\ \citenamefont
  {Sasorov}(2008)}]{PhysRevE.78.060103}%
  \BibitemOpen
  \bibfield  {author} {\bibinfo {author} {\bibfnamefont {B.}~\bibnamefont
  {Meerson}}\ and\ \bibinfo {author} {\bibfnamefont {P.~V.}\ \bibnamefont
  {Sasorov}},\ }\href {https://doi.org/10.1103/PhysRevE.78.060103} {\bibfield
  {journal} {\bibinfo  {journal} {Phys. Rev. E}\ }\textbf {\bibinfo {volume}
  {78}},\ \bibinfo {pages} {060103} (\bibinfo {year} {2008})}\BibitemShut
  {NoStop}%
\bibitem [{\citenamefont {Huang}\ \emph {et~al.}(2015)\citenamefont {Huang},
  \citenamefont {Hauert},\ and\ \citenamefont
  {Traulsen}}]{doi:10.1073/pnas.1418745112}%
  \BibitemOpen
  \bibfield  {author} {\bibinfo {author} {\bibfnamefont {W.}~\bibnamefont
  {Huang}}, \bibinfo {author} {\bibfnamefont {C.}~\bibnamefont {Hauert}},\ and\
  \bibinfo {author} {\bibfnamefont {A.}~\bibnamefont {Traulsen}},\ }\href
  {https://doi.org/10.1073/pnas.1418745112} {\bibfield  {journal} {\bibinfo
  {journal} {Proceedings of the National Academy of Sciences}\ }\textbf
  {\bibinfo {volume} {112}},\ \bibinfo {pages} {9064} (\bibinfo {year}
  {2015})}\BibitemShut {NoStop}%
\bibitem [{\citenamefont {Bellac}(2011)}]{Bellac:2011kqa}%
  \BibitemOpen
  \bibfield  {author} {\bibinfo {author} {\bibfnamefont {M.~L.}\ \bibnamefont
  {Bellac}},\ }\href {https://doi.org/10.1017/CBO9780511721700} {\emph
  {\bibinfo {title} {{Thermal Field Theory}}}},\ Cambridge Monographs on
  Mathematical Physics\ (\bibinfo  {publisher} {Cambridge University Press},\
  \bibinfo {year} {2011})\BibitemShut {NoStop}%
\bibitem [{\citenamefont {Kapusta}\ and\ \citenamefont
  {Gale}(2023)}]{Kapusta:2023eix}%
  \BibitemOpen
  \bibfield  {author} {\bibinfo {author} {\bibfnamefont {J.~I.}\ \bibnamefont
  {Kapusta}}\ and\ \bibinfo {author} {\bibfnamefont {C.}~\bibnamefont {Gale}},\
  }\href {https://doi.org/10.1017/9781009401968} {\emph {\bibinfo {title}
  {{Finite-Temperature Field Theory}}}},\ Cambridge Monographs on Mathematical
  Physics\ (\bibinfo  {publisher} {Cambridge University Press},\ \bibinfo
  {year} {2023})\BibitemShut {NoStop}%
\end{thebibliography}%

\end{document}